\documentclass[
 aps,
 pra,
 showpacs,
 superscriptaddress,
 amsmath,amssymb,nofootinbib,
reprint,
 twocolumn
]{revtex4-1}          % twocolumn

\RequirePackage[T1]{fontenc}

\usepackage[english]{babel}
\usepackage[T1]{fontenc}
\usepackage{graphicx}
\usepackage{subfigure}
\usepackage{amsmath}
\usepackage{amssymb}
\usepackage{slashed}
\usepackage{dsfont}
\usepackage{lipsum}

\begin{document}

\title{O(2)-scaling in finite and infinite volume}

\author{Paul Springer}
  \email{paul.springer@mytum.de}
  \affiliation{Physik Department, Technische Universit\"{a}t M\"{u}nchen, D-85747 Garching, Germany}
  \author{Bertram Klein}
  \email{bklein@tum.de}
  \affiliation{Physik Department, Technische Universit\"{a}t M\"{u}nchen, D-85747 Garching, Germany}

\date{\today}

\begin{abstract}

The exact nature of the chiral crossover in QCD is still under investigation. In $N_f=2$ and $N_f=(2+1)$ lattice simulations with staggered fermions the expected O($N$)-scaling behavior was observed. However, it is still not clear whether this behavior falls into the O(2) or O(4) universality class. To resolve this issue, a careful scaling and finite-size scaling analysis of the lattice results are needed. We use a functional renormalization group to perform a new investigation of the finite-size scaling regions in O(2)- and O(4)-models. We also investigate the behavior of the critical fluctuations by means of the $4^{\text{th}}$-order Binder cumulant. The finite-size analysis of this quantity provides an additional way for determining the universality class of the chiral transition in lattice QCD.
\end{abstract}

\maketitle

\section{Introduction} \label{sec:Introduction}

Quantum Chromodynamics (QCD) at finite temperatures and chemical potential is currently one of the most actively researched topics in theoretical physics. In particular, the exact nature of the transition from hadronic phase to quark-gluon plasma is of great importance for interpretation of experimental results from heavy ion collisions \cite{BraunMunzinger:2003zd} and for the understanding of the evolution of the universe in its early stage.

Two transitions take place in QCD: the confinement-deconfinement and the chiral transition. The first one turns out to be a first order phase transition for pure gluonic systems, while it becomes a crossover in the presence of quarks. In contrast, the nature of the chiral transition depends on the number of dynamical quark flavors, but also on the strength of the explicit symmetry breaking and the strength of the chiral anomaly \cite{Pisarski:1983ms, Chandrasekharan:2007}. For zero chemical potential, vanishing anomaly and two massless quark flavors it is expected to be a phase transition of second order. In this case QCD falls into the O(4) universality class \cite{Pisarski:1983ms}. For massive quarks, however, the phase transition disappears and we expect to observe a crossover which should still fulfill the O(4) scaling behavior. In the presence of the strong chiral anomaly and for two quark flavors in the chiral limit one expects that the chiral phase transition is of first order \cite{Pisarski:1983ms, Chandrasekharan:2007}.

A convenient method to study full QCD are lattice simulations. One particular way to implement fermions on the lattice is to use staggered
fermions. This method allows to reproduce pion masses which are close to the chiral limit. However, staggered fermions exhibit only a reduced chiral
symmetry and we can also expect to observe a chiral transition governed by O(2)-symmetry. Finite simulation volumes poses an additional problem: continuous chiral symmetry cannot be broken spontaneously in a finite volume and explicit symmetry breaking in form of finite quark masses is mandatory. So, no real continuous phase transition can be observed in such simulations but only a crossover. These facts lead to complications in the interpretation of lattice data and hence, to ambiguous results: For a long time, data provided by lattice simulations with two dynamical or (2+1) quarks using the staggered formulation did not exhibit the expected O($N$)-scaling \cite{Aoki:1998wg, Bernard:1999fv, Engels:2005rr} or gave evidence for a first-order phase transition \cite{D'Elia:2005bv, Cossu:2007mn}. Results from present-day calculations are in a very good agreement with
O($N$)-scaling behavior \cite{Kogut:2006gt, Ejiri:2009ac, Bazavov:2011nk}. Nonetheless, there is still an ambiguity about the exact universality class of the transition observed.
However, especially the finite simulation volumes can affect the universal behavior of the thermodynamic observables in a very specific way, depending on the nature of the transition in the limit $V \to \infty$. Therefore, finite-volume effects could help to shed light on the nature of the chiral crossover observed in the lattice QCD \cite{Kogut:2006gt, Braun:2008sg}.

The chiral transition seems to be a continuous one. In this case long-range fluctuations are dominant and the microscopic details of the system are no longer relevant. The observed behavior is then basically determined by symmetries and dimensionality or, with other words, by the universality class of the system. It means that close to the transition point the thermodynamic quantities obey power-laws and scaling functions which are characterized by universal critical exponents. Even in the presence of finite volumes the behavior of the thermodynamic observables can be described by still universal finite-size scaling functions.

Since long-range fluctuations play such a prominent role at continuous transitions, we can try to investigate the nature of the chiral transition in lattice QCD using the higher-order fluctuations. For this purpose we can use a quantity called Binder cumulant \cite{Binder:1981sa}. The Binder cumulant has already been used
successfully for determining the critical value of the quark mass for three degenerated
massive flavors, where the phase transition is continuous and falls into the Z(2) universality
class \cite{Karsch:2001nf, deForcrand:2003hx}.

So far, scaling functions, finite-size scaling functions and Binder cumulant have been determined mainly by using O($N$)-symmetric spin-model lattice simulations \cite{Engels:2001bq, Ballesteros:1996bd, Engels:2000xw, STMAZ.01038989, Kanaya:1994qe}. These results were often used in the analysis of lattice QCD data \cite{Kogut:2006gt, Engels:2005rr, Engels:2001bq, Mendes:2006zf, Mendes:2007ve, Engels:2005}. In the present paper we use an alternative technical framework, the functional renormalization group approach (FRG). We calculate scaling and finite-size scaling functions for the 3-dimensional linear O(2) model and compare our results with findings from \cite{Braun:2007td} and \cite{Braun:2008sg} where the O(4) model was considered. For both models we also calculate the Binder cumulant and investigate its behavior in finite volumes. Our approach is complementing O($N$)-spin-model lattice simulations. Our results can be directly applied in the scaling analysis of the lattice QCD data.

This paper is organized as follows: In Sec.~\ref{sec:MethodAndModel} we briefly discuss the setup of our FRG formalism and introduce scaling and finite-size scaling functions. Then in Sec.~\ref{sec:BinderTheory} we define Binder cumulant of the $4^{\text{th}}$ order and discuss its properties. After, in Sec.~\ref{sec:ResultsInfiniteV} and Sec.~\ref{sec:ResultsFiniteV} we present our results for critical exponents and scaling functions in infinite and finite volumes. In Sec.~\ref{sec:ResultsBinder} we show and discuss numerical results for the Binder cumulant. Our concluding remarks can be found in Sec.~\ref{sec:Conclusion}.

\section{Method and Model} \label{sec:MethodAndModel}

\subsection{FRG applied to O($N$)-models}

In our investigations we use the functional renormalization group applied to continuous O($N$)-symmetric models in 3 spatial dimensions. At the ultra-violate (UV) scale $\Lambda$ the bare action of these models is defined as
\begin{equation}
\Gamma_{\Lambda}[\boldsymbol{\phi}]=\int \text{d}^{d}x\Big(\frac{1}{2}\tilde{Z}_{\boldsymbol{\phi}}(\partial_{\mu} \boldsymbol{\phi})^2+U_{\Lambda}(\boldsymbol{\phi}^{\text{ }2})\Big)\text{ ,}
\label{eq:GammaInPositionSpace}
\end{equation}
where $\boldsymbol{\phi} = (\sigma, \pi_1, \dots, \pi_{N-1})^T \in \mathbb{R}^{N}$ represents multiple, effective scalar degrees of freedom, $\tilde{Z}_{\phi}$ is a wave function renormalization and $U_{\Lambda}(\boldsymbol{\phi}\,^2)$ an effective potential defined at the scale $\Lambda$.  This potential should depend only on the powers of $\boldsymbol{\phi}\,^2$ and should exhibit O($N$)-symmetry. If we introduce the finite explicit symmetry breaking in our calculations, we add the term $-H \sigma$ to the potential. In our calculations we use the so-called local potential approximation (LPA) where we assume that expectation values of the fields do not have any spatial dependence and the wave function renormalization is constant: $\tilde{Z}_{\phi}=1$. This assumption leads to vanishing anomalous dimension, $\eta = 0$. Even though the value of $\eta$ for O($N$)-models was measured to be finite, it is still relatively small compared to one, see, e.g., Ref.~\cite{Tetradis:1993ts}. Therefore our assumption of constant wave function renormalization is well justified.

The RG flow of the effective action can be described by the Wetterich flow equation \cite{Wetterich:1992yh}:
\begin{equation}
\partial_t \Gamma_k = \frac{1}{2} \text{Tr} \Big[ \partial_t R_k \Big(\Gamma_k^{(2)} + R_k  \Big)^{-1}\Big] \text{ ,}
\label{eq:Wetterich}
\end{equation}
where $t=\ln (k/ \Lambda)$. The function $R_k$ in the expression above is the so-called regulator function which controls the Wilsonian momentum-shell integrations and has to fulfill certain constraints \cite{Wetterich:1992yh}. We are free in the choice of $R_k$, so we can use it for the optimization of the flow \cite{Litim:2000ci, Litim:2001fd, Litim:2001up, Pawlowski:2005xe}. In this work we use the 3-dimensional Litim's optimized regulator \cite{Litim:2001up} given by
\begin{align}
R_k(p^2)=(k^2-\boldsymbol{p}^{\text{ }2}) \theta(k^2-\boldsymbol{p}^{\text{ }2})\text{ .}
\end{align}
Using this regulator function and applying the LPA we find the following flow equation of $U_{k}$ for the case of infinitely large volumes:
\begin{equation}
\partial_t U_k[\phi^2] =\frac{k^5}{6 \pi^2}\Big[\frac{1}{k^2+M_{\sigma}^2 } + \frac{(N-1)}{k^2+M_{\pi}^2 }\Big]\text{ .}
\label{eq:RGFlowForOurCalculation}
\end{equation}
The scale-dependent expressions $M_{\sigma}^2$ and $M_{\pi}^2$ are defined as the eigenvalues of the second-derivative matrix of $U_k$ and are equal to the square of bosonic masses in the limit $k \to 0$. Note that they still depend on the field $\phi$.

Since we do not know the exact form of the effective potential $U_{k}$, we expand it around its particular minimum  $\boldsymbol{\phi}=(\sigma_0 (k),0, \dots, 0)^T$:
\begin{equation}
U_{k}=\sum \limits_{m=0}^M a_m(k)(\boldsymbol{\phi}^{\text{ }2}-\sigma_0^2(k))^m-H\cdot\sigma\text{ ,}
\label{eq:PotentialExpansionBroken}
\end{equation}
where $\sigma_0 (k) \neq 0$ only if the symmetry is broken. If we also introduce a finite explicit symmetry-breaking, we choose $H \neq 0$, otherwise $H=0$

Expansion of the right-hand side of RG-flow, Eq.~\eqref{eq:RGFlowForOurCalculation}, around the minimum of $U_k$ will provide us with a set of highly coupled differential equations for couplings $a_m(k)$. In order to fix the scale-dependent expectation value of field $\sigma$, we use an additional condition which ensures that we are expanding our potential around the actual physical minimum:
\begin{equation}
\frac{\partial U_k}{\partial \sigma}\Big|_{\sigma=\sigma_0(k),\boldsymbol{\pi}^{2}=0}=0 \quad \Rightarrow \quad 2 a_1(k) \sigma_0(k)=H\text{ .}
\end{equation}
Now we have all information we need to solve the RG flow. However, the number of allowed couplings $a_m(k)$ is infinite. So we have to choose an appropriate truncation scheme for our potential. Unfortunately there is no argument allowing us to neglect couplings of higher order a priori like it can be done in perturbative calculations. However, in \cite{Tetradis:1993ts,Litim:2001hk} it was found that inclusion of a small number of couplings is already sufficient for O($N$)-models. In our present calculations we have used $M=6$ and have checked explicitly that our results do not change considerably if we use additional couplings.

At the Wetterich flow equation for finite volume, we replace spatial momenta integrations by a sum over discrete momenta. In this step it is possible to use periodic or anti-periodic boundary condition for discretized momenta. Since most of lattice QCD simulations use periodic boundary condition, we also use this:
\begin{equation}
	\int\limits_{-\infty}^{\infty}dp_i \rightarrow  \frac{2 \pi}{L} \sum\limits_{n_i \in  \mathbb{Z}}\text{ ,} \qquad \text{ with } i=1,2,3
\end{equation}
with
\begin{equation}
\quad \boldsymbol{p}\,^2=\frac{4\pi^2}{L^2}\sum\limits_{i=1}^{3}n_i^2 \text{ .}
\end{equation}
The flow of our effective potential \eqref{eq:RGFlowForOurCalculation} is then modified to
\begin{equation}
\partial_t U_k[\phi^2]=k^5\Big[\frac{1}{k^2+M_{\sigma}^2 } + \frac{(N-1)}{k^2+M_{\pi}^2 }\Big] \mathcal{B}(kL)\text{ ,}
\label{eq:RGFlowForOurCalculationFiniteVolume}
\end{equation}
where $\mathcal{B}(kL)$ is a mode-counting function which includes all information we need for fluctuation modes allowed in a particular volume with extent $L$. It is provided by
\begin{equation}
\mathcal{B}(kL) = \frac{1}{(kL)^3}\sum \limits_{\boldsymbol{n} \in \mathcal{Z}^3} \Theta((kL)^2-\boldsymbol{p}^{\text{ 2}}L^2)\text{ .}
\label{eq:ModeCountingFunction}
\end{equation}
The limiting behavior of this function is important for understanding the contributions of different fluctuation modes to the RG-flow: Since we use periodic boundary condition in our calculations, the mode counting function behaves in the limit $kL \to 0$ as:
\begin{equation}
\lim \limits_{kL \to 0} \mathcal{B}(kL) \sim \frac{1}{(kL)^3}\text{ .}
\end{equation}
This is due to existence of zero-momentum mode for the choice of periodic boundary condition. For very small volumes the dynamics of the system are basically governed by this zero-momentum mode.

For infinitely large volumes the sum in Eq.~\eqref{eq:ModeCountingFunction} becomes the volume of a sphere with the radius $r=kL/(2\pi)$ and we find
\begin{equation}
\lim \limits_{kL \to \infty} \mathcal{B}(kL)=\frac{1}{6 \pi^2}\text{ .}
\end{equation}
Thus, we recover the flow equation for infinite volume. In our finite-volume calculations we use the same expansion of the effective potential as given in Eq.~\eqref{eq:PotentialExpansionBroken}. Since for finite volumes a finite explicit symmetry-breaking term is mandatory, we have to choose $H \neq 0$.

The FRG approach described above allows us to calculate scaling and finite-size scaling functions as well as the Binder cumulant. These functions are introduced below.

\subsection{Scaling functions in infinite volume} \label{sec:ScalingInfinite}

The behavior in the vicinity of the critical transition point is governed by the free-energy density. This quantity consists of a singular and a regular part
\begin{equation}
	f=f_s(T,H)+f_r(T,H)\text{ ,}
\end{equation}
where only the singular part is responsible for critical behavior, whereas the regular one leads at most to some finite corrections.

In our model we assume that only temperature $T$ and explicit symmetry breaking $H$ are the relevant couplings. In order to remove all system-specific scales, we introduce the rescaled temperature $t$ and the rescaled symmetry-breaking field $h$ as new variables:
\begin{align}
		t = \frac{T-T_C}{T_0} \text{ ,}	 \qquad \qquad h = \frac{H}{H_0} \text{ ,}
	\label{eq:RescaledVariables}
\end{align}
where $T_C$ is the critical temperature and $T_0$ and $H_0$ are system-dependent normalization constants. Then, the singular part of the free-energy density is a function of $t$ and $h$: $f_s(T,H)=f_s(t,h)$.

Close to the critical point, correlation length has infinite range $(\xi \to \infty)$ and the system becomes scale invariant. Therefore the singular part of the free energy density is invariant under a rescaling of the length with a factor $a$:
\begin{equation}
f_s(t,h)=a^{-d} f_s(a^{y_t}t,a^{y_h}h)\text{ .}
\label{eq:FreeEnergyDensityRescaled}
\end{equation}
The critical exponents can be then expressed in terms of $y_t$ and $y_h$:
\begin{align}
	y_t = \frac{1}{\nu}\text{ ,} \qquad \qquad y_h = \frac{\beta \delta}{\nu}\text{ .}
\end{align}
Since we have two relevant couplings in our model, only two critical exponents are independent. All others can be obtained using the following scaling laws:
\begin{align}
	\begin{split}
	\gamma=(2-\eta)\nu \text{ , }& \enspace \gamma=\beta(\delta-1) \text{ , } \enspace\beta=\frac{1}{2}(d-2+\eta)\nu \text{ ,}\\
	\nu d&=\beta(1+\delta) \text{ , }\enspace \delta=\frac{d+2-\eta}{d-2+\eta}\text{ .}
	\end{split}
	\label{eq:ScalingLaws}
\end{align}

The critical exponents $\beta$ and $\delta$ govern the behavior of the order parameter M, which is associated in our model with the expectation value of the field $\sigma$:
\begin{align}
	M(t,h=0) = (-t)^{\beta}\text{ ,} \qquad \qquad M(t=0,h) = h^{1/\delta}\text{ .}
	\label{eq:OrderParameterPowerLaws}
\end{align}
The critical exponents $\nu$ and $\gamma$ describe the behavior of the correlation length $\xi$ and longitudinal susceptibility $\chi$ correspondingly:
\begin{align}
	\xi \propto |t|^{-\nu}\text{ ,} \qquad \qquad \chi\propto |t|^{-\gamma}\text{ .}
	\label{eq:PowerLawXi}
\end{align}

In this paper we investigate only the behavior of the order parameter $M$. Using the rescaled form of the free energy density, Eq.~\ref{eq:FreeEnergyDensityRescaled}, $M$ can be derived from its thermodynamic definition:
\begin{align}
	\begin{split}
	M &= \sigma_0 = -\frac{\partial f_s}{\partial H}= h^{1/ \delta} f_M(z)\text{ ,}
	\end{split}
	\label{eq:ScalingFunctInf}
\end{align}
where $z=t/h^{1/(\beta \delta)}$ is a new single scaling variable and is invariant under the rescaling of $t$ and $h$.

In the expression above, $f_M(z)$ is a scaling function for the order parameter $M$. It turns out that this function is universal for an universality class. For other thermodynamic observables the corresponding scaling functions can be derived at the similar manner.

\subsection{Scaling functions in finite volume}

If a thermodynamic system is put into a finite volume, the correlation length is bounded from above by the system extent $L$. Since the critical point can be reached only in the limit $L \to \infty$, the system extent becomes an additional coupling, and the critical behavior changes. Since the singular part of the free-energy density is now a function of three variables, we need to introduce two scaling variables in order to define a finite-size scaling function.

According to Fisher's finite-size scaling hypothesis \cite{Fisher:1971ks}, the ratio of thermodynamic quantities in the finite-size system to those in an infinite system depends on only the ratio $\xi/L$. This implies that in the absence of $h$ the system extent has to scale with $t$ exactly in the same way as $\xi$, Eq.~\eqref{eq:PowerLawXi}. We also know from definition of $z$ that $t \propto h^{1/(\beta \delta)}$. These observations allow us to introduce a new scaling variable $h^*=h l^{\beta \delta /\nu}$, where $l=L/L_0$ is the renormalized system extent, and $L_0$ is system specific. Then, using Eq.~\eqref{eq:ScalingFunctInf} for the order parameter, we find 
\begin{equation}
M(t,h,l)=l^{-\beta/\nu}Q_M(z,h^*) \text{ ,}
\label{eq:ScalingFunctFinite}
\end{equation}
where $Q(z,h^*)$ is a finite-size scaling function in leading order. For completeness, we also specify the leading-order finite-size scaling correction \cite{Fisher:1971ks}:
\begin{equation}
M(t,h,L)=l^{-\beta/\nu}\left[Q_M(z,h^*)+\frac{1}{l^{\omega}} \tilde{Q}_M^{(1)}(z,h^*)+ \ldots\right] \text{ ,}
\label{eq:FiniteSizeScalingCorrections}
\end{equation}
where $\omega$ is the critical exponent associated with the first irrelevant operator in the renormalization group flow. This additional term influences the behavior of the system in the vicinity of the critical point for small $L$ and needs to be removed in order to isolate the universal finite-size scaling function. In fact, we observe this correction in our calculations for small volumes. However, since we use very large volumes to fix $Q(z,h^*)$, we will neglect finite-size scaling corrections and use the above leading-order expression for the order parameter.

\section{Binder cumulant} \label{sec:BinderTheory}

In addition to the universal scaling function for the order parameter we also investigate a higher-order fluctuation quantity, the so-called Binder cumulant of the $4^{\text{th}}$ order \cite{Binder:1981sa}. In the infinite-volume limit, this quantity exhibits a value at the critical point which is specific for some particular universality class. Therefore, it is often used for the localization of the critical point, e.g. \cite{Karsch:2001nf, deForcrand:2003hx, Cucchieri:2002hu}.

The universal values of the Binder cumulant for the three-dimensional O(2)- and O(4)-models have already been determined at high accuracy by using spin-model lattice simulations \cite{STMAZ.01038989, Kanaya:1994qe}. To our knowledge, the Binder cumulant has never been calculated previously using FRG. In \ref{BinderCumulant} we present such a calculation for the O(2) and O(4) universality classes in LPA. Using our approach we can also investigate the influence of finite-volume effects on the Binder cumulant.

The convenient definition of the $4^{\text{th}}$-order Binder cumulant is given by
\begin{align}
B_4=\frac{\langle(\boldsymbol{\phi}\,^2)^2 \rangle}{\langle \boldsymbol{\phi}\,^2 \rangle ^2} \text{ .}
\label{eq:BinderDefinition}
\end{align}
with $\boldsymbol{\phi}=(\sigma, \pi_i)^{\text{T}}$ and $i$ number of Goldstone modes. In this quantity we compare the total contribution of all possible fluctuations of the 4$^{\text{th}}$ order to the contribution from trivial Gaussian fluctuations.

In the phase with broken symmetry, the system obtains a very high stability. In this regime fluctuations are suppressed by powers of $\frac{1}{V}$ (see \ref{BinderCumulant}) and we expect
\begin{align}
\langle (\boldsymbol{\phi}\,^2)^2 \rangle \to M ^4\text{ ,} \qquad \qquad \langle \boldsymbol{\phi}\,^2 \rangle \to M ^2\text{ ,}
\end{align}
for $T \ll T_C$. Therefore, for any O($N$)-model the value of $B_4$ should approach $1$ for decreasing temperatures.

In contrast, for $T \gg T_C$ fluctuations become the leading contributions. At large $T$ the contribution from fluctuations of some particular order goes into saturation and $B_4$ approaches some finite value. For models with a different number of degrees of freedom, the number of fluctuations contributing to $\langle \boldsymbol{\phi}^4 \rangle$ is obviously not identical and the Binder cumulant $B_4$ approaches a value, which is specific for a particular model. Our calculations suggest a general expression for this limiting behavior for O($N$)-symmetric models:
\begin{equation}
B_4=\frac{N+2}{N}\text{ .}
\end{equation}
So, for $T \gg T_C$ we expect $B_4=2$ for O(2)- and $B_4=3/2$ for O(4)-model. These values are confirmed by our numerical calculations.

Additionally the Binder cumulant is directly a finite-size-scaling function. If we consider almost vanishing symmetry-breaking field $H$, this function is given by
\begin{align}
B_4=Q_B(t L^{1/\nu}, L^{-\omega}, \dots) \text{ .}
\end{align}
$Q_B$ depends on the scaling variable $t L^{1/\nu}$ and possibly on some other irrelevant operators which we specify only up to the leading order, i.e., we assume only the finite-size corrections proportional to $L^{-\omega}$, with $\omega > 0$. Expanding this finite-size scaling function to the lowest obtainable order in both variables
\begin{equation}
Q_B(t L^{1/\nu}, L^{-\omega})=a_0 +a_1 t L^{1/\nu} + a_2 L^{-\omega} + \ldots \text{ .}
\label{eq:ExpansionOfBinder}
\end{equation}
From this expression we see that exactly at the critical point ($t=0$, $L \to \infty$) the Binder cumulant is simply given by a constant. This constant is in general different for different universality classes. Therefore it can be potentially applied in order to determine universality class of a particular system, e.g., in lattice QCD simulations.

\begin{table*}
\centering
\begin{tabular*}{\textwidth}{@{\extracolsep{\fill}}lrrrrl@{}}
\hline
\hline
  O(2) &  \multicolumn{2}{c}{$H=0$} &   \multicolumn{2}{c}{$H= 1.0 \times 10^{-13} \enspace \text{MeV}^{5/2}$}&   \\
   & derivative & fit & derivative & fit & final \\
   \hline
  $\beta$ & 0.3537 & 0.3536  & 0.3540 & 0.3538 & 0.3538(2)  \\
  $\gamma$ & 1.4143 & 1.4139 & 1.4134 & 1.4142 & 1.4140(4)  \\
\hline
\hline
 & & & & &\\
 & & & & &\\
\hline
\hline

 O(2)  & derivative & fit & $\gamma / \beta+1$ & $(d+2-\eta)/(d-2+\eta)$ & final \\
   \hline
  $\delta$ & 4.9997 & 4.9995  & 4.9966 & 5.0000 & 4.9990(15)  \\
\hline
\hline
\end{tabular*}
\caption{\small{Our results for the critical exponents $\beta$, $\gamma$ and $\delta$ for the O(2)-model in LPA. Using different fit techniques allows us to estimate the uncertainty of our determinations. In the evaluation of $\delta$ we also use our results for $\beta$ and $\gamma$ and theoretical prediction for $\delta$. Since in our formalism $\eta=0$ and $d=3$, the theoretical prediction for $\delta$ is $5.0$.}}
\label{tab:CriticalExponents1}
\end{table*}

\section{Infinite-volume scaling} \label{sec:ResultsInfiniteV}

\subsection{Critical exponents}

To this day, critical exponents for three-dimensional O(2)-model were calculated at very high accuracy using different methods such as  lattice Monte-Carlo simulations of spins \cite{Ballesteros:1996bd, Engels:2000xw}, perturbative field-theoretical \cite{Guida:1998bx, Butera:1997ak}, and RG calculations \cite{Litim:2001hk, Tetradis:1993ts, Bohr:2000gp, VonGersdorff:2000kp}. Since we use LPA in this particular work which implies the vanishing anomalous dimension $\eta=0$, we do not aim to add it to this list. However, in order to provide a consistent evaluation of scaling functions, we need to calculate critical exponents within the LPA and to use them in the following analysis.

In our formalism we use $d=3$, therefore, we cannot define temperature in the field-theoretical sense. Nonetheless, we can find a parameter in the RG-flow which does the same job as $T$, i.e., controls the phase transition. In our case it is the initial value of the expectation value of $\phi$ at the UV scale $\Lambda$. So, we suppose the existence of an expansion $(\phi_0(\Lambda)-\phi_0^{\text{critical}}(\Lambda)) \sim (T-T_C)$ \cite{PhysRevLett.77.873,Adams:1995cv,Braun:2007td}.

In our calculations we use $\Lambda=10$ GeV. For this setup we obtain the following critical UV-value:
\begin{equation}
	\phi_0^{\text{critical}}(\Lambda) =37.996488987996596 \text{ MeV}^{1/2} \text{ .}
\end{equation}
As a point of fact, this accuracy in $\phi_0^{\text{critical}}(\Lambda)$ is necessary in order to observe the fixed point in the RG-flow. For simplicity we will use in the following the notation $T$ and $T_C$ for $\phi_0(\Lambda)$ and $\phi_0^{\text{critical}}(\Lambda)$.

We calculate the critical exponents $\beta$ and $\gamma$ in the chiral limit ($H=0$) using the power-laws discussed in Sec.~\ref{sec:ScalingInfinite}. We use two different fitting techniques: direct fitting and taking the numerical derivative of $\log(M)$ and $\log(\chi)$ in the limit $t\to0$. In order to estimate error, we repeat these calculations for small explicit symmetry breaking \mbox{$H=10^{-13} \text{ MeV}^{5/2}$}. Our results can be found in the upper part of Table \ref{tab:CriticalExponents1}. In the following analysis we use averaged values of the critical exponents.

For the determination of $\delta$ we evaluate the order parameter $M$ exactly at the critical temperature for different symmetry-breaking fields $H$. Once again, we provide a direct fit of our results and a fit to the numerical derivative of $\log(M)$. In addition, we calculate $\delta$ using the scaling law $\delta=\gamma/\beta+1$ and our results for $\beta$ and $\gamma$ from Table \ref{tab:CriticalExponents1}. In the determination of the averaged value of $\delta$, we also use the theoretical prediction:
\begin{equation}
	\delta = \frac{d+2-\eta}{d-2+\eta}\text{ .}
\end{equation}
In $d=3$ dimensions and in the absence of the anomalous dimension (LPA) we expect $\delta=5$. Our critical exponents are summarized in the lower part of Table \ref{tab:CriticalExponents1}. 

Our results seem to be consistent and agree within 0.1\% with the results from \cite{Litim:2001hk}, where exact renormalization group in LPA was used to calculate  the critical exponent $\nu=2\beta$ (in LPA). Never the less, we should keep in mind that all our calculations include some additional systematic errors corresponding to truncation of the effective potential $U(\phi^2)$ and to neglecting the kinetic terms of higher order in the effective action $\Gamma[\phi]$. However, we have considered a relatively large number of $n$-point couplings ($n_{max}=12$) and results in Refs.~\cite{Tetradis:1993ts, Litim:2001hk} show that inclusion of a small number of couplings is already sufficient for calculations in O($N$)-models. Therefore, we estimate that systematic uncertainty of our results due to the truncation of $U(\phi^2)$ is comparatively small.

Once we have determined the values of the critical exponents, we can calculate the normalization constants $T_0$ and $H_0$. The logarithms of these constants appear in additive terms in $\log M$ at $H=0$ and at $T=T_C$ respectively, Eq.~\eqref{eq:OrderParameterPowerLaws}. We obtain
\begin{align}
	\begin{split}
	T_0 &=0.0046874(8)\text{ MeV$^{1/2}$},\\
	H_0 &=13.837(13) \text{MeV$^{5/2}$ .}
	\end{split}
\end{align}
Estimating the error of $T_0$, we perform calculations with the additional small explicit symmetry-breaking field: $H=10^{-13}$ and $10^{-12} \text{ MeV}^{5/2}$. In analogy, for $H_0$ we repeat our calculations in the presence of non-vanishing but small $(T-T_C)=1 \times 10^{-13}$ and $2 \times 10^{-13} \text{ MeV}^{1/2}$.

In our calculations in finite volumes an additional system-specific scale $L$ should be removed in order to determine finite-size scaling functions \cite{Braun:2008sg}. A possible normalization choice is given by
\begin{align}
\xi = L_0 |t|^{-\nu} \text{ .}
\end{align}
And we find
\begin{align}
	\begin{split}
	L_0 &=62.206(1) \text{ fm.}
	\end{split}
\end{align}
The error is estimated using the same explicit symmetry-breaking fields as in the case of error estimation for $T_0$.

\subsection{Scaling functions}

\begin{figure*}[t!]
\begin{minipage}[hbt]{\columnwidth}
	\centering
	\includegraphics[width=1\columnwidth]{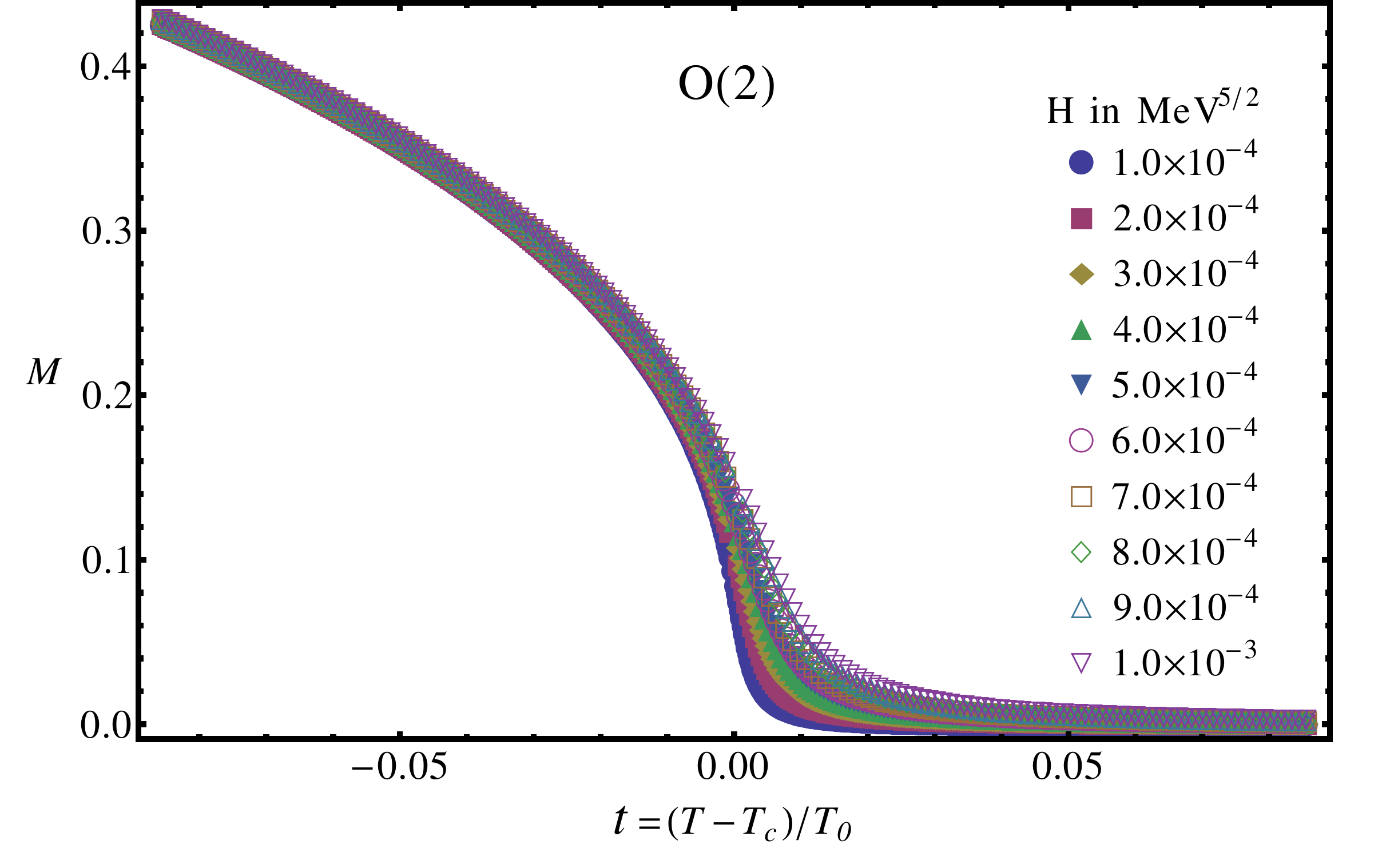}
\end{minipage}
\begin{minipage}[hbt]{\columnwidth}
	\centering
	\includegraphics[width=1\columnwidth]{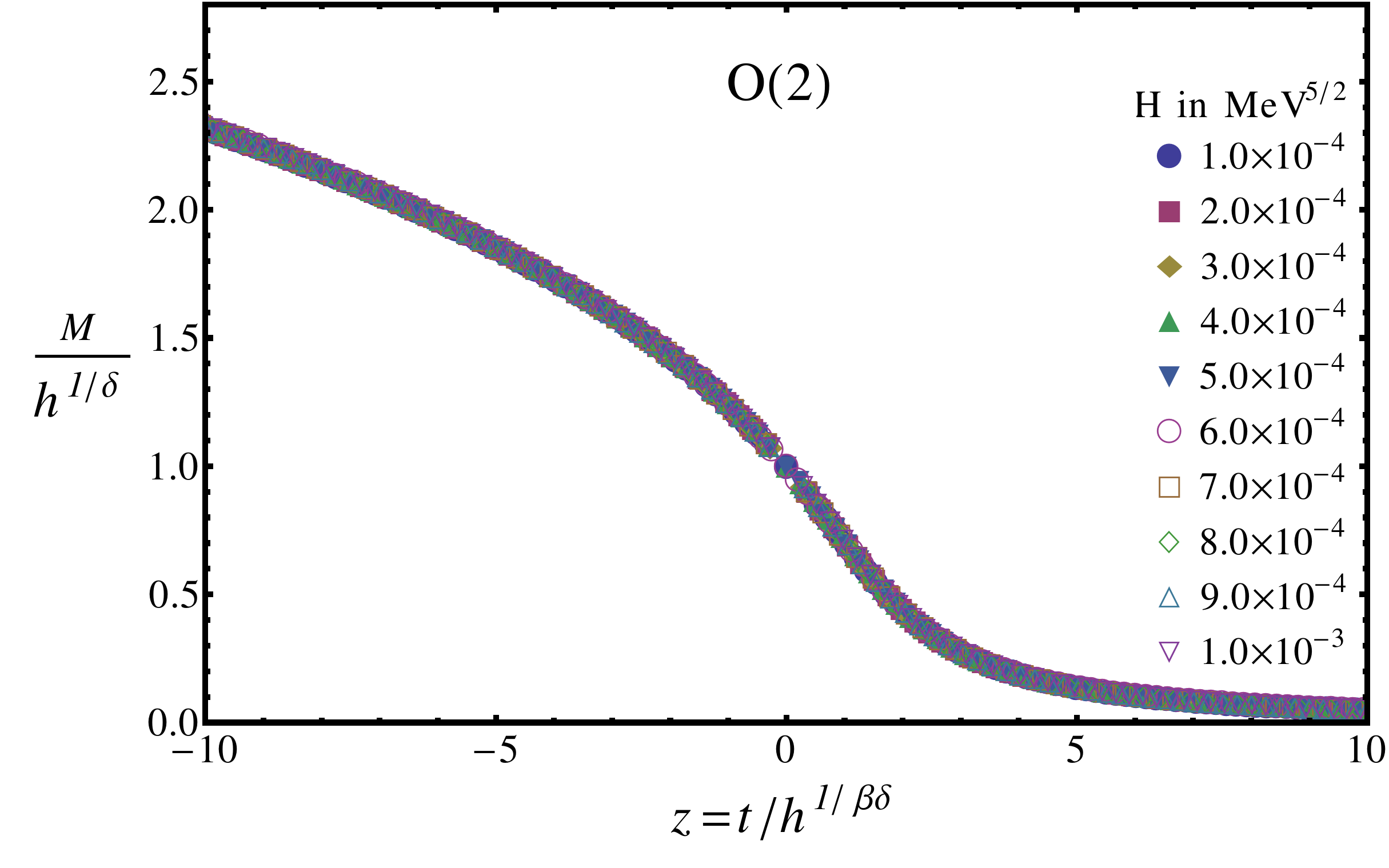}
\end{minipage}
\caption{\small{In this figure we present our results for the order parameter in O(2)-model calculated in the presence of the very small symmetry-breaking field. In the \textit{left} part we show the unrescaled, in the \textit{right} part the rescaled data. The rescaled data fall onto one line, i.e., they exhibit the ideal universal scaling behavior.}}
\label{fig:OrderParameterDataFieldsO2}
\end{figure*}

We calculate the order parameter over a wide range of values for $T$ and in the presence of some small symmetry-breaking $H$. Our results are presented in the left part of the Fig.~\ref{fig:OrderParameterDataFieldsO2}. If we consider a fixed temperature interval, we observe that for smaller values of $H$ the order parameter in the vicinity of $T_C$ decreases more rapidly and asymptotically approaches zero already at very small temperature $t$. It means that for small symmetry-breaking fields the magnetization behaves approximately as for $H=0$ if we are sufficiently far away from the critical temperature. However, with increasing $H$ the crossover character becomes more distinct.

We rescale the data for $M$ using the critical exponents and normalization constants determined in the previous subsection. Our results are shown in the right part of Fig.~\ref{fig:OrderParameterDataFieldsO2}. The rescaled data falls perfectly into one line, i.e., we observe ideal scaling behavior.

\begin{figure}[t]
\centering
\includegraphics[width=1\columnwidth]{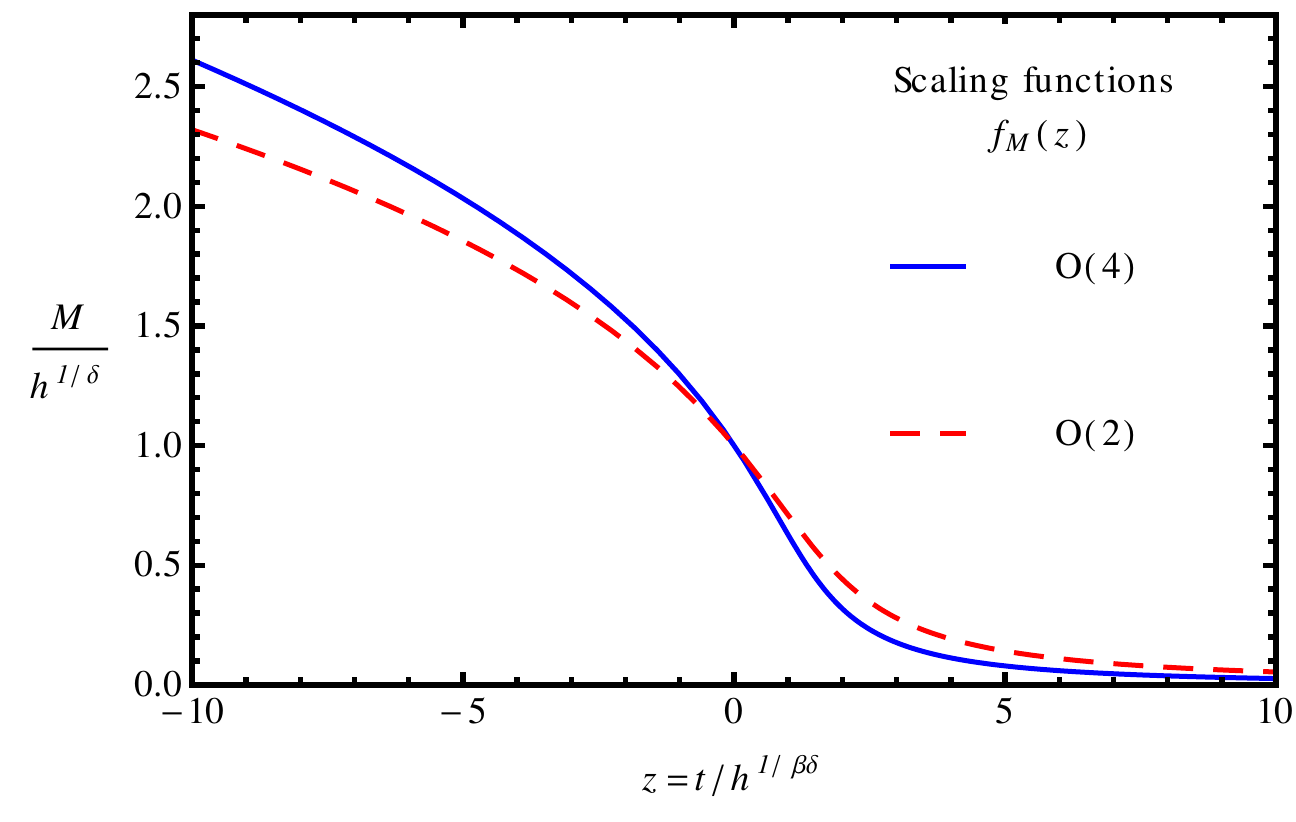}
\caption{\small{Our results for O(2)- and O(4)-scaling functions $f_M(z)$. Both these functions behave in a very similar way, whereas in the vicinity of $T_C$ the scaling function for the O(4)-model decreases a bit faster with increasing $z$ than that for the O(2)-model.}}
\label{fig:ScalefunctionsOrderParameterTogether}
\end{figure}

For small $H$ the scaling corrections are negligible. Therefore, we use the data for the smallest $H$ to determine the scaling function for the order parameter. Our result for $f_M(z)$ is presented in Fig.~\ref{fig:ScalefunctionsOrderParameterTogether}. In this figure we also plot the order parameter scaling function for the O(4)-model which we obtain, using the same formalism as in the O(2) case (see also~\cite{Braun:2007td}). We observe that these two functions are very similar. This similarity is a reason why many investigations of scaling properties of lattice QCD, where the lattice data is fitted to the scaling functions of the order parameter, led to ambiguous results \cite{Bernard:1999fv, Engels:2005rr, Ejiri:2009ac}.

\begin{figure}[t]
\centering
\includegraphics[width=1\columnwidth]{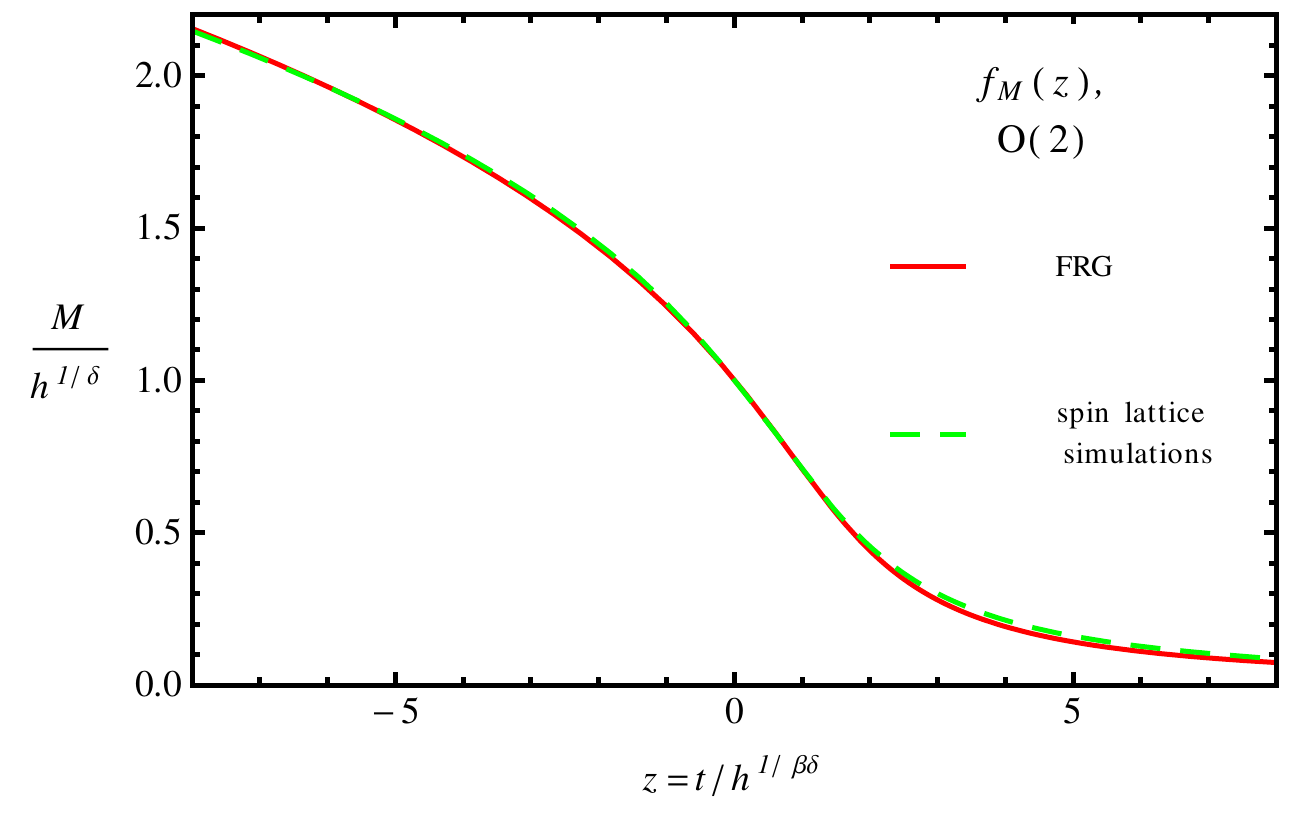}
\caption{\small{Scaling function of the order parameter $M$ for three dimensional O(2)-model calculated within LPA using critical exponents from Tab.~\ref{tab:CriticalExponents1} is compared with results from a lattice spin simulation which already include anomalous dimension \cite{Engels:2001bq}. We observe an excellent agreement.}}
\label{fig:InfiniteVolumeScalingFunctionsForTheOrderParameter}
\end{figure}

As a check of our results, we compare our scaling function in the infinite-volume limit with corresponding function obtained using lattice spin simulations of O(2)-model \cite{Engels:2001bq}, Fig.~\ref{fig:InfiniteVolumeScalingFunctionsForTheOrderParameter}. Though the calculations in \cite{Engels:2001bq} already include the non-vanishing anomalous dimension $\eta$, we see an almost perfect agreement with our result. Therefore, we conclude that our formalism provides reasonable data. We infer that the RG-approach is a very appropriate tool for the determination of the scaling behavior.

\section{Finite-volume scaling} \label{sec:ResultsFiniteV}

\begin{figure*}[t]
\centering
\includegraphics[width=1\columnwidth]{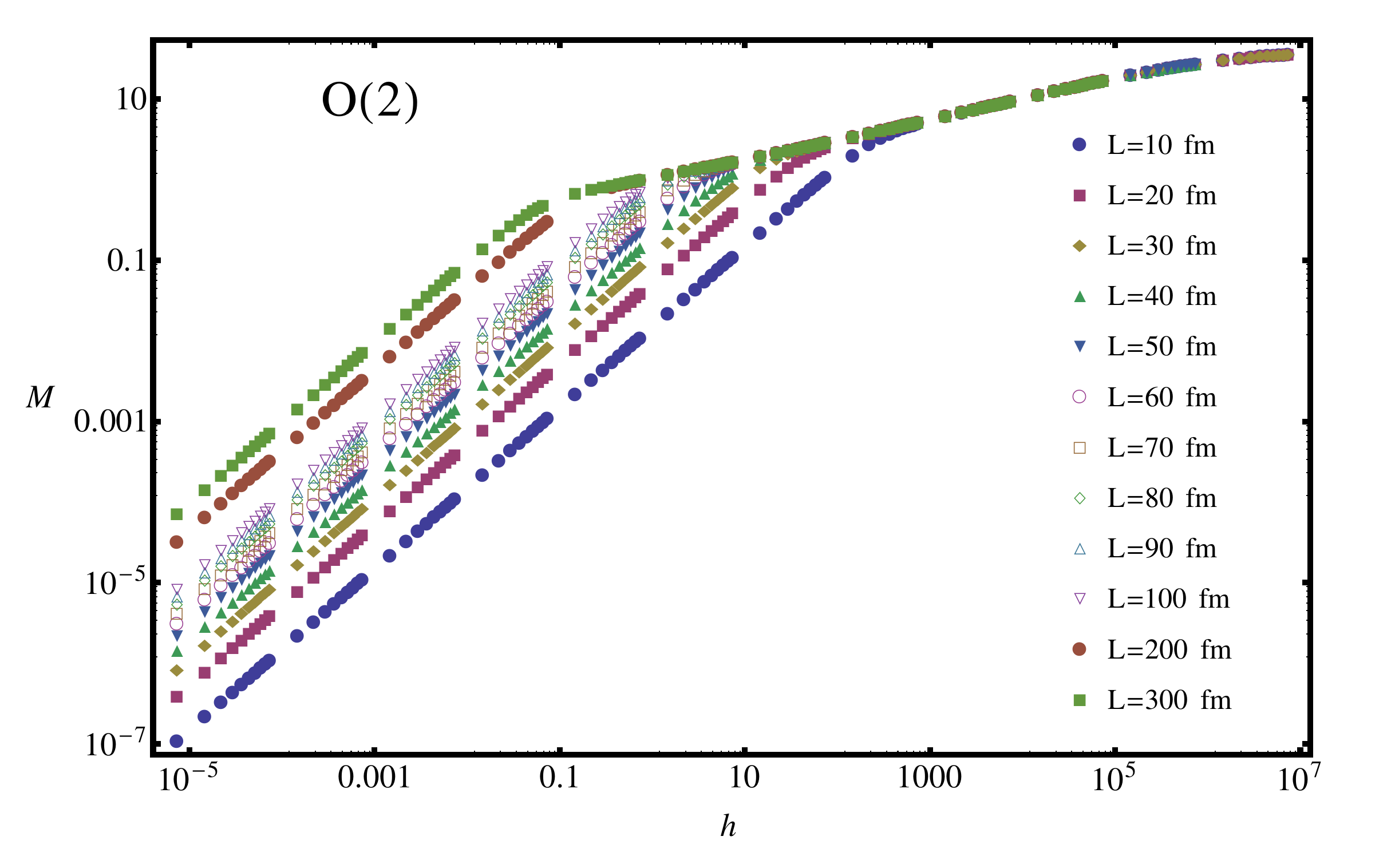}
\includegraphics[width=1\columnwidth]{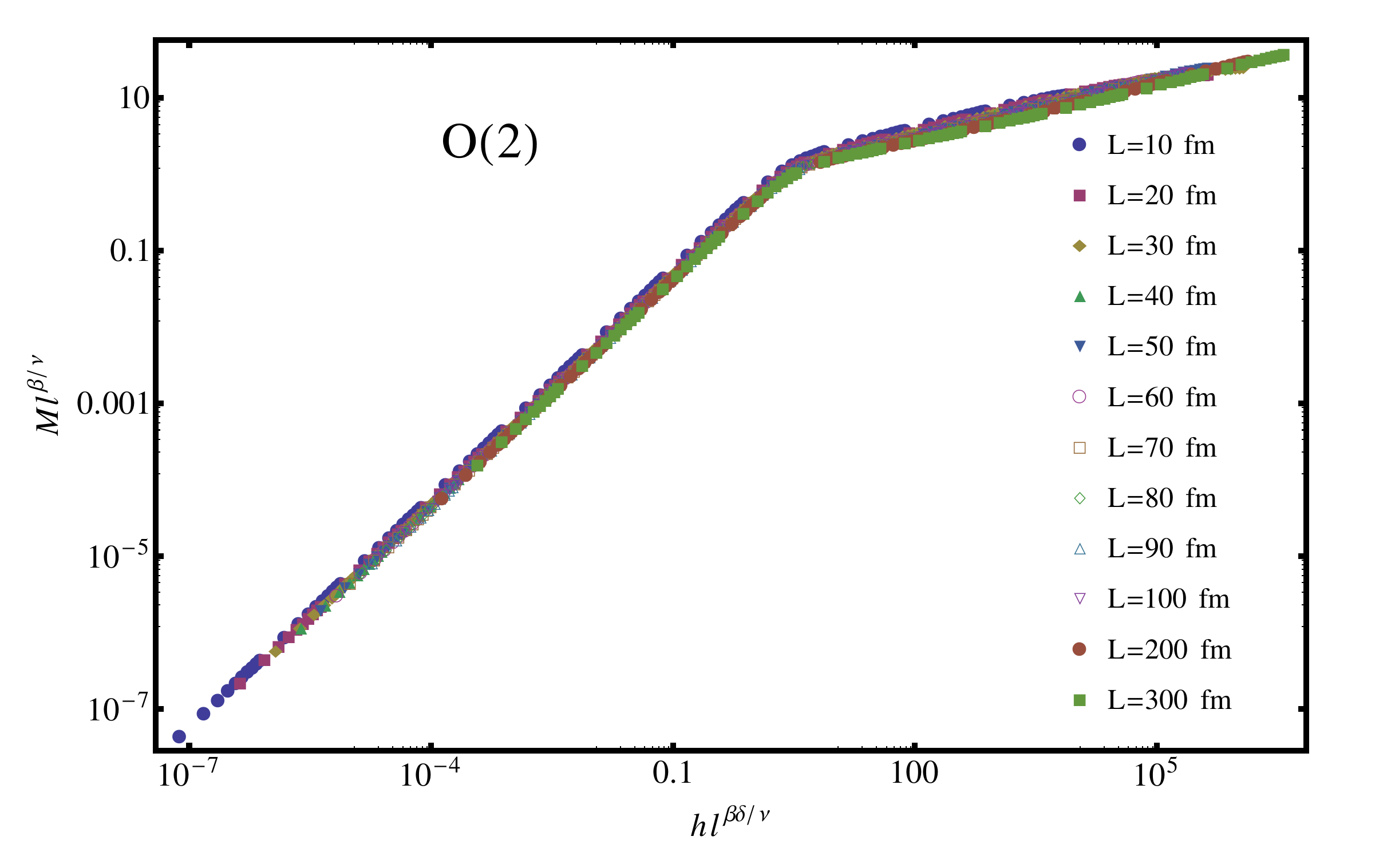}
\caption{\small{Our results for the order parameter $M$ in different finite volumes $L=10-300 \text{ fm}$ for three-dimensional O(2)-model calculated within LPA at $z=0$. In the \textit{left} part we present the unrescaled, in the \textit{right} part the rescaled data. We see two different scaling regions: For large $h$, we see the scaling behavior valid for the limit $L \to \infty$, for small $h$ we observe finite-size scaling. In the rescaled plot we observe that curves for different volumes fall almost perfectly onto one line. For small volumes we observe some deviations from the universal finite-size scaling behavior due to the non-universal scaling corrections. However, they are negligible for very large volumes.}}
\label{fig:FiniteVolumeOrderParameter}
\end{figure*}

We determine the finite-volume scaling function for the order parameter. A similar investigation of the finite-size scaling in the O(4)-model was already provided in \cite{Braun:2008sg}. We calculate the order parameter $M$ as a function of the symmetry-breaking field $h$ for different finite-volume sizes, $L=$10-300 fm exactly at $T=T_C$. Our results are shown in the double-logarithmic representation in the left part of Fig.~\ref{fig:FiniteVolumeOrderParameter}. In this plot we can distinguish two different regions in the behavior of $M$: For very large symmetry-breaking fields we observe the same quantitative behavior for all volumes which we have considered. The slope of the curves here is very close to $1/\delta$, i.e, we observe approximately the same power law as in the limit $L\to\infty$, Eq.~\eqref{eq:OrderParameterPowerLaws}. This behavior appears because of the large masses of the fluctuations. In this situation, the correlation length $\xi$ is much smaller than the extent of the system and therefore, finite $L$ does not influence the critical behavior. If however, the external symmetry-breaking field $h$ becomes smaller, the mass of fluctuations decreases and the correlation length grows. At some point $\xi$ becomes so large that its value becomes comparable with $L$. Therefore, the infinite-volume scaling behavior becomes affected by the extent of the system and we observe the finite-size scaling region.

\begin{figure}[t]
\centering
\includegraphics[width=1\columnwidth]{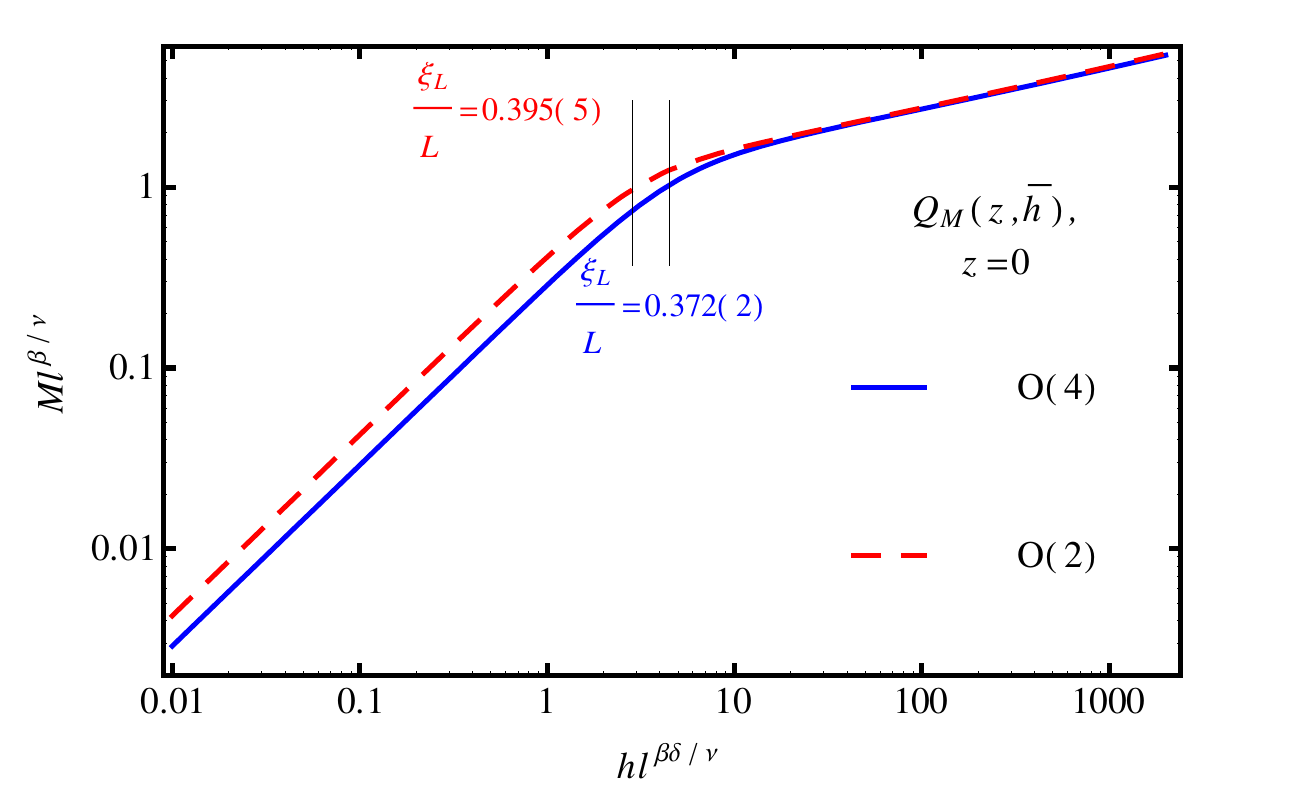}
\caption{\small{Finite-size scaling functions for the order parameter $M$ for three dimensional O(2)- and O(4)-models exactly at the critical point $t=0$ are plotted in the double-logarithmic representation as functions of the dimensionless scaling variable $hl^{\beta \delta / \nu}$. The finite-size scaling regions are labeled by the corresponding universal values of $\xi_L/L$. For O(2) this value is given by $\xi_L/L=0.395(5)$, for O(4) by $\xi_L/L=0.372(2)$.}}
\label{fig:FiniteVolumeScalingFunctionsForTheOrderParameter}
\end{figure}

We rescale our results and consider the rescaled order parameter $Ml^{\beta/\nu}$ as a function of the dimensionless scaling variable $hl^{\beta \delta / \nu}$, right part of Fig.~\ref{fig:FiniteVolumeOrderParameter}. We observe that curves for different volumes fall almost perfectly into one line. However, the agreement becomes worse for decreasing $L$. This fact is explained by the presence of non-universal finite-size corrections. Since these corrections scale with the system extent as $L^{-\omega}$, they are negligible for very large volumes. On account of this, we determine the finite-size scaling function by using the data for the largest volume we have calculated ($L=300 \text{fm}$). In Fig.~\ref{fig:FiniteVolumeScalingFunctionsForTheOrderParameter} we present our O(2)-finite-volume scaling function together with corresponding scaling function for the O(4)-model.

From the perspective of the analysis of lattice QCD results, an interesting feature of the finite-size scaling function is the region where the universal finite-size scaling behavior appears: we need to know it, in order to decide, where simulations need to be done in parameter space in order to use finite- or infinite-size scaling behavior in the analysis of lattice results. Also, the finite size scaling regions for the O(2)- and O(4)- models can be different and, probably, can be used to distinguish O(2) and O(4) universality classes. Since this regime in scaling behavior arises if the correlation length is in the order of the system extent, it is self-evident to use the universal dimensionless combination $\xi/L$ to describe the change in the scaling behavior. We use the longitudinal part of the correlation length $\xi_L=1/m_{\sigma}$ as a measure for $\xi$. As also shown in the Fig.~\ref{fig:FiniteVolumeScalingFunctionsForTheOrderParameter}, we determine the value of this quantity, at the point where the finite-size scaling appears. It is given by
\begin{align}
	\begin{split}
	\frac{\xi_L}{L} &=0.395(5) \text{ for O(2),}\\	
	\frac{\xi_L}{L} &=0.372(2) \text{ for O(4).}\\
	\end{split}
\end{align}
In order to determine these values, we find a linear parametrization for our data in a double-logarithmic representation for large and small $h$. We use the value of $h$ at the point of intersection of these two lines as the point where the influence of $L$ becomes dominant. To estimate the errors we determine the finite-size scaling region using our results for somewhat smaller volumes: $L=100 \text{, }200 \text{ fm}$.

While the description of the finite-size scaling regions by the universal value of $\xi_L/L$ is self-evident, it is inapplicable for practical purposes since the correlation length is difficult to measure in lattice simulations. Therefore we look for an alternative description. Since the correlation length of fluctuations is bounded from above by the inverse pion mass $m_{\pi}$, the finite-size scaling should appear if the wave-length of the pion is
of the order of the system extent:
\begin{equation}
	\frac{1}{m_{\pi}} \sim L.
\end{equation}
Therefore, we can use the universal dimensionless quantity $m_{\pi} L$ for the description of the point where the finite-size scaling occurs. Since the pion mass and the system extent are used as inputs in lattice simulations, this way to estimate the finite-size scaling region is easier to apply in lattice QCD. Using our approach we find:
\begin{align}
	\begin{split}
	(m_{\pi}L) &=2.01(1) \text{ for O(2),}\\
	(m_{\pi}L) &=2.12(2) \text{ for O(4).}\\
	\end{split}
\end{align}
Here we use the same calculation and error estimation technique as for $\xi_L/L$. 

The values of $\xi_L/L$ and $m_{\pi} L$ are very similar for O(2)- and O(4)-models. Therefore, using the finite-size scaling regions to determine the nature of the chiral transition in lattice QCD seems to be difficult. Never the less, these results can be used in the scaling analysis of the lattice QCD data in order to clarify in which scaling region some particular data set should be located. Assuming O(2) or O(4) scaling behavior for lattice QCD, we should fit the data to the finite-size scaling functions if we are in the region $m_{\pi}L \lesssim 2$. On the other hand, for $m_{\pi}L \gtrsim 2$ we can expect only infinite-size scaling.

\section{Binder cumulants}\label{sec:ResultsBinder}

\begin{figure}[t]
\centering
\includegraphics[width=1\columnwidth]{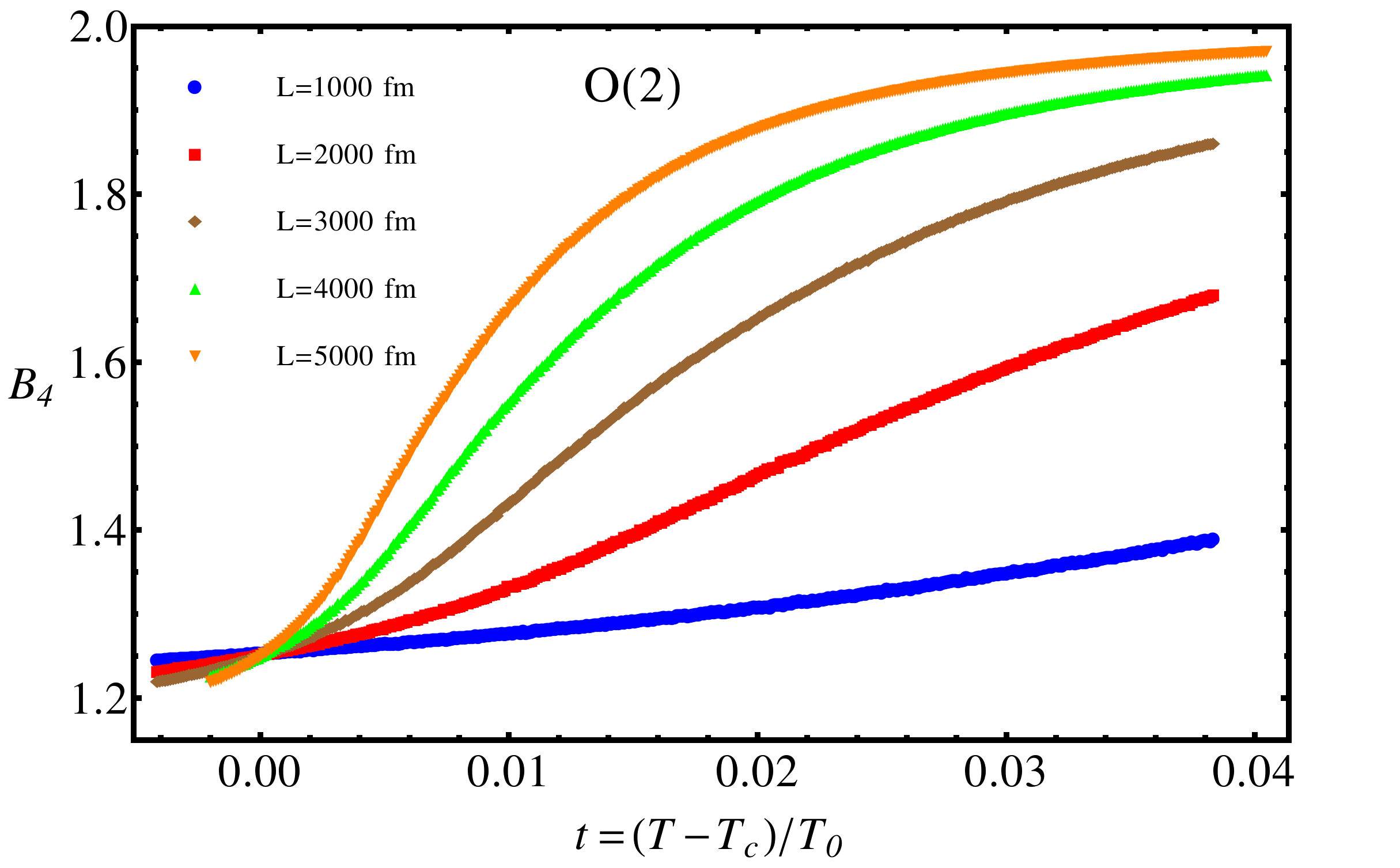}
\caption{\small{Results for the Binder cumulant $B_4$ for the O(2)-model at almost vanishing symmetry-breaking field ($H=10^{-13} \text{ MeV}^{5/2}$) as a function of temperature. The results for the largest volumes we have investigated are shown: $L \in \{1000 \text{ fm, } 2000 \text{ fm, } 3000 \text{ fm, } 4000 \text{ fm, } 5000 \text{ fm} \}$. We observe the correct asymptotic behavior at very large temperatures, $B_4 \to 2$. The behavior in the limit $T \ll T_C$, $B_4 \to 1$, has also been checked. Considered on the scale of this plot, it seems that graphs for different volumes cross at $T=T_C$.}}
\label{fig:BinderCumulant}
\end{figure}

Next, we calculate $B_4$ as a function for the temperature $t$. Since the Binder cumulant is defined in our approach for finite volumes, we cannot use zero symmetry-breaking. However, we can still employ an almost vanishing symmetry-breaking field $H=10^{-13} \text{ Mev}^{5/2}$. We use very different volumes, $L=10-5000 \text{ fm}$. In Fig.~\ref{fig:BinderCumulant} we present our O(2)-results for five largest volumes considered. We have checked the correct asymptotic behavior for small $T$, $B_4 \to 1$, for all volume sizes. As expected, for large temperatures $B_4$ approaches $2$ for the O(2)-model. We observe the slope of $B_4$ decreases with decreasing $L$. For smallest volumes we have considered the corresponding limit can be achieved only at the temperatures far beyond critical one. In this figure, we can see that graphs for different volumes cross at nearly one and the same point, close to the critical temperature. This crossing point corresponds to the universal value of the Binder cumulant in the thermodynamic limit. However, since we use large, but still finite volumes, our results include some finite-size corrections. We exclude them according to a method described in \cite{Binder:1981sa, Cucchieri:2002hu, Ballesteros:1996bd}. Our calculation provide following results:
\begin{align}
\begin{split}
B_4 &=1.2491(39) \text{ for the O(2)-model ,}\\
B_4 &=1.0836(10) \text{ for the O(4)-model ,}
\end{split}
\end{align}
where errors are estimated using different spatial extends \linebreak $L \in \{1000 \text{ fm, } 2000 \text{ fm, } 3000 \text{ fm, } 4000 \text{ fm, } 5000 \text{ fm} \}$.

Our value for the O(2)-models is very close to the value obtained in \cite{Cucchieri:2002hu}: $B_4 =1.242(2)$. In the case of O(4), we observe a somewhat larger deviation from the value determined using spin-model lattice simulations \cite{Kanaya:1994qe}: $B_4 =1.092(3)$. In both cases the discrepancy between our RG-results and results from Monte Carlo simulations is smaller than 1$\%$. This deviation is explained by the fact that we neglect anomalous dimension $\eta$. However and as a matter of fact, the discrepancy is very small. Therefore, once again we can infer that the influence of finite $\eta$ on the scaling behavior is almost negligible.

\begin{figure}[t]
\centering
\includegraphics[width=1\columnwidth]{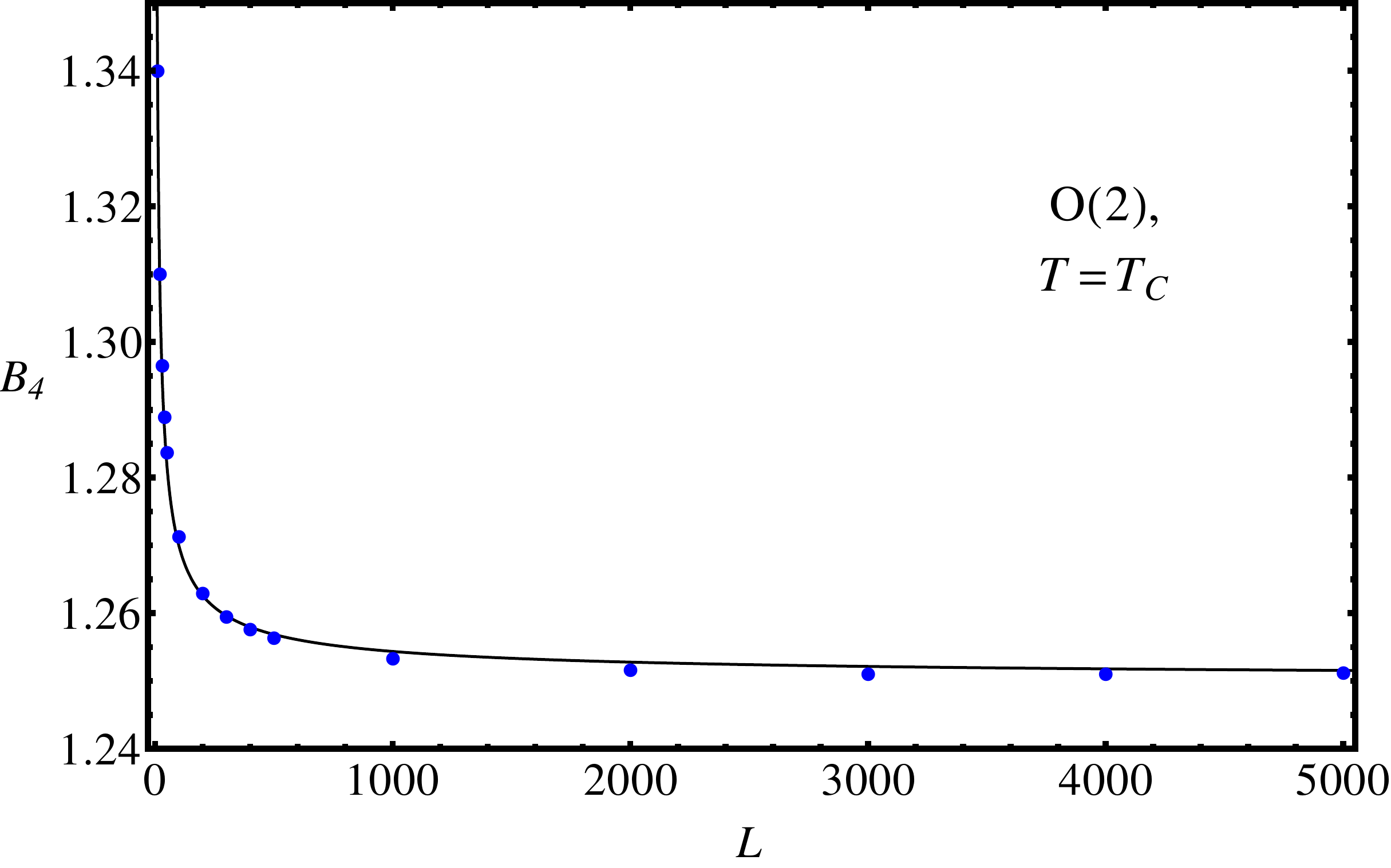}
\caption{\small{Here we show the values of $B_4$ at $tL^{1/\nu}=0$ for O(2)-model for all volumes we have considered, $L = 10 - 5000 \text{ fm}$. We observe that for volumes of some $10 \text{ fm}$ the non-universal finite-size corrections to the value of the Binder cumulant at $T=T_C$ become noticeable. We also provide fits of our results to the expansion given in Eq.~\eqref{eq:ExpansionOfBinder}. This form agrees very well with our data points.}}
\label{fig:BinderCumulantRescaledCrossingPoints}
\end{figure}

We investigate the influence of the finite volume on the value of $B_4$ at the critical temperature. In Fig.~\ref{fig:BinderCumulantRescaledCrossingPoints} we present our results for the O(2)-model. We include results from data sets for all volumes we have considered. We observe that for volumes which are typical for lattice QCD simulations ($\sim 10-30 \text{ fm}$), the finite-size corrections to the universal value of $B_4$ are in the order of $3-8 \%$ for both models. We have also checked that the finite-size correction is very well described by the leading order expansion, Eq.~\eqref{eq:ExpansionOfBinder}. The corresponding fit for the O(2)-model is also presented in Fig.~\ref{fig:BinderCumulantRescaledCrossingPoints} (for $\omega$ we use the value $\omega=0.6712$ obtained in \cite{Litim:2001hk}).

\begin{figure}[t]
\centering
\includegraphics[width=1.09\columnwidth]{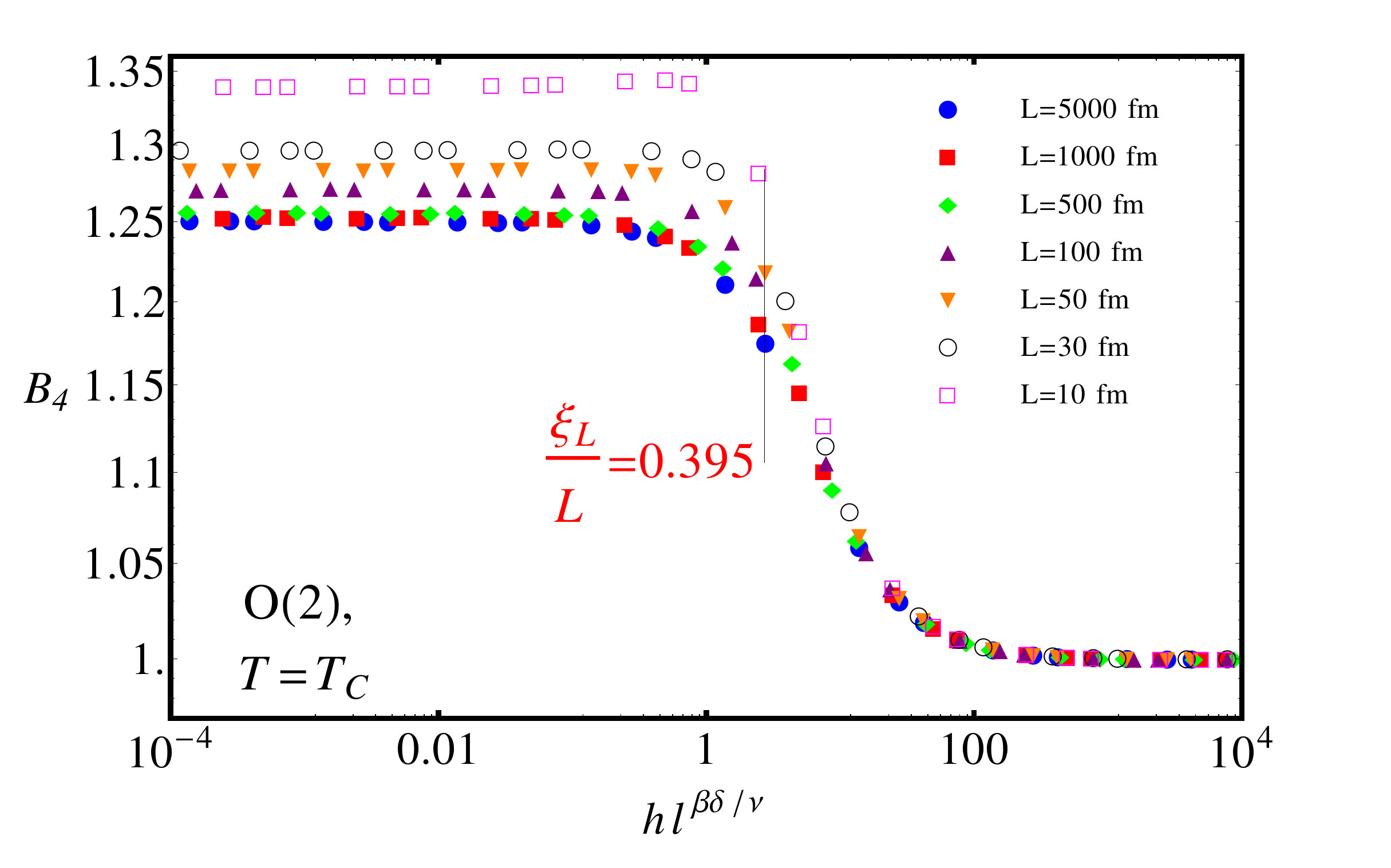}
\caption{\small{The rescaled results for $B_4(T_C)$ as a function of the symmetry-breaking field for O(2)-model. We observe that the data points do not fall onto one line. We can clearly see that deviations are mostly caused by non-universal finite-size corrections. We also present in this plot the value of $\xi_L/L$ which correspond to the finite-size scaling region of the order parameter. We observe that this value describes the finite-size scaling region of $B_4$ very well.}}
\label{fig:BinderCumulantDifferentVolumesHVariationAtTCRescaled}
\end{figure}
\begin{figure}[t]
\centering
\includegraphics[width=1.09\columnwidth]{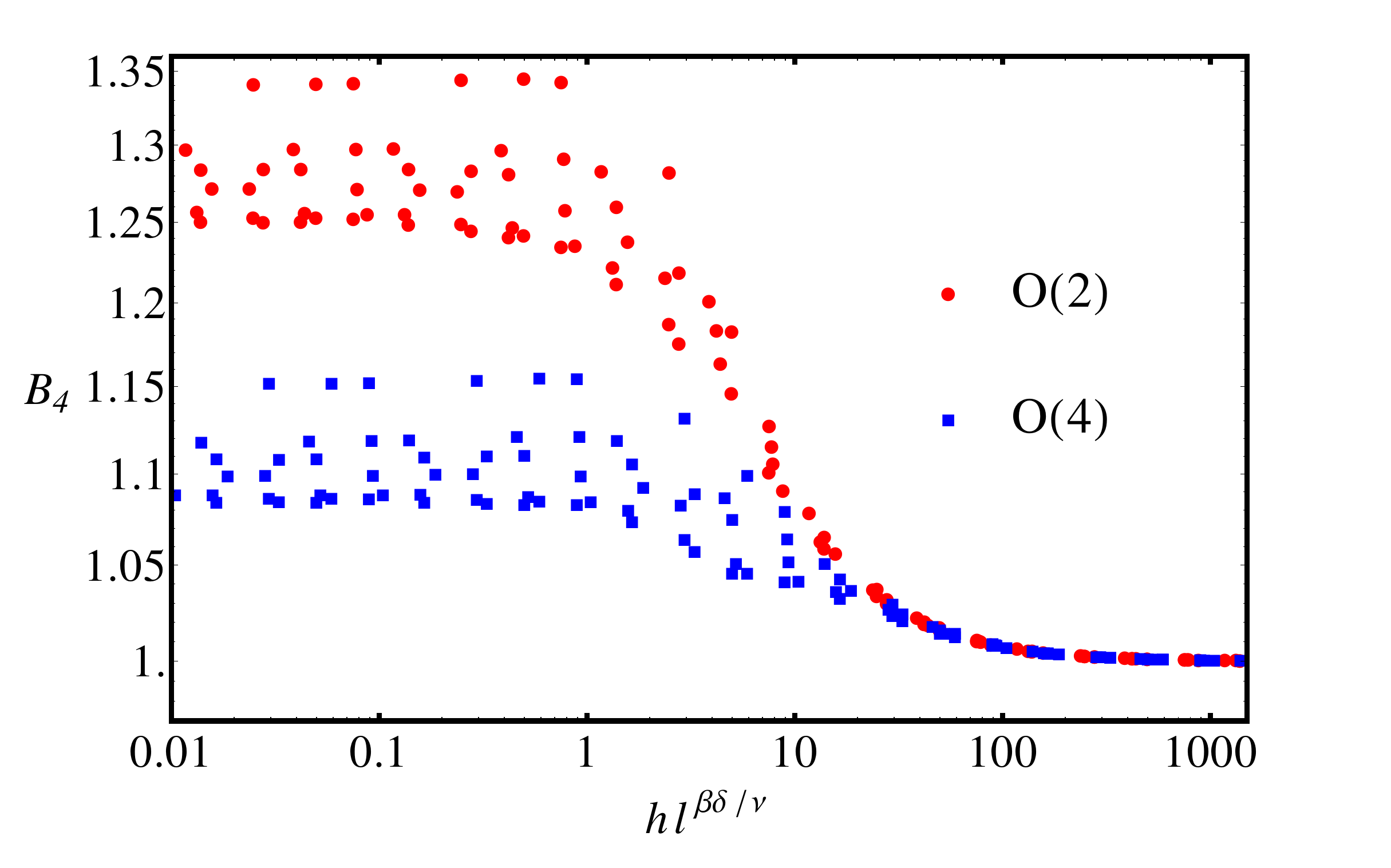}
\caption{\small{Here we present the rescaled data as in the Fig.~\ref{fig:BinderCumulantDifferentVolumesHVariationAtTCRescaled} but now for the O(2)- \textit{(red points)}  and for the O(4)-model \textit{(blue squares)}. For both models points with the largest values of $B_4$ correspond to calculations with $L=10 \text{ fm}$ and points with the smallest values of $B_4$ to calculations with $L=5000 \text{ fm}$. We observe that even in the presence of non-universal finite-size corrections we can clearly distinguish between O(2)- and O(4)-models.}}
\label{fig:BinderCumulantDifferentVolumesHVariationAtTCO2ANDO4RescaledAllData}
\end{figure}

In addition, we investigate the behavior of the universal value of $B_4(T_C)$ as a function of the symmetry-breaking field $h$. In Fig.~\ref{fig:BinderCumulantDifferentVolumesHVariationAtTCRescaled} we present our results for some selected volumes in rescaled form. For small $h$ we observe the same finite-size effects as discussed in Fig.~\ref{fig:BinderCumulantRescaledCrossingPoints}. We do not observe any noticeable finite-mass corrections. For all volumes we find the value of $B_4(T_C)$ to remain constant for small $h$. However, at some point the value of the Binder cumulant decreases very fast and approaches 1. These two regimes are finite-size and infinite-size scaling regions of $B_4(T_C)$: For small symmetry-breaking fields the correlation length is in the order of the system extent $L$ and the finite-volume effects become dominant. However, if we consider very large symmetry-breaking fields, the correlation length decreases, the finite-size effects become less pronounced, and the system behaves approximately in the same way as in infinitely large volume. With increasing $h$ we are moving away from the critical point and the system becomes more and more orderly. In the orderly phase, however, the Binder cumulant should approach the limit 1.

In Fig.~\ref{fig:BinderCumulantDifferentVolumesHVariationAtTCRescaled} we also plot our results for the values of $\xi_L/L$ at the onset of the finite-size scaling regions. We find that our results from Fig.~\ref{fig:FiniteVolumeScalingFunctionsForTheOrderParameter} are in very good agreement with the finite-size scaling regions which we observe for the Binder cumulant.

In Fig.~\ref{fig:BinderCumulantDifferentVolumesHVariationAtTCO2ANDO4RescaledAllData} we present our results for O(2)- and O(4)-models together in one plot.
We observe that even in the presence of the finite-volume corrections, the regions where we can expect to measure $B_4(T_C)$ in finite-volume lattice simulations do not overlap for O(2)- and O(4)-models. Of course, the smallest volume we have considered is $L=10 \text{ fm}$, and finite-size corrections should be larger for even smaller volumes. However, the typical lattice QCD simulations are performed in volumes with $L \sim 10 \text{ fm}$. Also we cannot exclude a possibility that non-universal finite-size corrections in the lattice QCD simulations are even larger than in our approach.  However, we see in the Fig.~\ref{fig:BinderCumulantDifferentVolumesHVariationAtTCO2ANDO4RescaledAllData} that the gap between O(2) and O(4) results is relatively large. It is at least of the order of the finite-size corrections we expect for $L=10 \text{ fm}$. Thus, we conclude that the universal value of the Binder cumulant is a very reasonable tool to distinguish O(2) and O(4) universality classes even in the presence of finite-size corrections to scaling. Therefore, $B_4(T_C)$ seems to be a very promising candidate for a criterion for determination of the universality class of the chiral transition in $N_f=2$ or $N_f=(2+1)$ lattice QCD. In order to use this result, lattice QCD simulations should be performed in the finite-size scaling region, i.e., at the values of $m_{\pi}L \lesssim 2$.

\section{Conclusions} \label{sec:Conclusion}

We have investigated the critical behavior in the continuous $\phi$-model with O(2)-symmetry in both infinite and finite volumes. In addition we have performed the same analysis for the O(4) case as it was already done in \cite{Braun:2007td, Braun:2008sg}. For these purposes we have applied functional RG approach in the local potential approximation. We have investigated regions where the finite-size scaling becomes dominant. We have also studied the behavior of the critical fluctuations in the vicinity of the critical point by means of the $4^{\text{th}}$-order Binder cumulant $B_4$. For this purpose we have derived an expression for the Binder cumulant in the context of FRG. It allows us to investigate higher-order fluctuations and to better understand the universal and limiting behavior of $B_4$ from a theoretical point of view. Though in the present work we consider only O(2)- and O(4)-models, our theoretical calculations can be applied to any O($N$)-symmetric model. We can also extend this method to Binder cumulants of higher order. In the case of two or $(2+1)$-flavors, the lattice QCD simulations cannot reach exactly the critical point because of finite $m_q$ and finite volumes. Therefore, additional analysis of the Binder cumulant as a function of the symmetry-breaking field and as a function of the system extent was needed in order to apply $B_4$ for the analysis of lattice QCD data.

We have considered the case of infinitely large volume in order to determine universal critical exponents valid in LPA. Our results are in a perfect agreement with results from \cite{Litim:2001hk}. This fact implies that possible systematic uncertainty caused by the specific truncation for the order of the potential which we used is very small. However, we still cannot exclude uncertainties arising from neglecting the higher order kinetic terms in the ansatz for the scale-dependent effective action.

We have also calculated universal scaling functions for the order parameter valid in infinitely large volumes. Our findings are in a very good agreement with results from \cite{Engels:2001bq}, where the authors have used lattice simulations for the three-dimensional O(2) spin-model. The tiny deviations we observed arise basically because of the anomalous dimension $\eta$ which we have neglected in our calculations.

In our finite-volume calculations we were able to determine the finite-size scaling function for the order parameter for O(2)-model exactly at the critical temperature. We have also described the regions where the finite-volume effects become dominant for O(2)- and O(4)-models using the universal values of $\xi_L/L=1/(m_{\sigma}L)$ and $m_{\pi}L$. We have found that finite-size scaling regions for these two models are similar. Therefore, the difference in the finite-size scaling regions for O(2)- and O(4)-models can probably not be used in order to determine the universality class of the chiral transition in $N_f=2$ or $N_f=(2+1)$ lattice QCD. Never the less, the finite-size scaling regions we have explicitly determined in this work are still useful for scaling analysis of lattice QCD data: We can use these results in order to decide whether a particular set of the simulation data should exhibit infinite-size or finite-size scaling behavior. The value of $m_{\pi}L$ which separates these two regimes is given for both models by $m_{\pi}L \approx 2$.

We have investigated the behavior of critical fluctuations close to the critical point by means of the Binder cumulant of the $4^{\text{th}}$ order, $B_4$. Our numerical calculations have confirmed our theoretical predictions about the limiting behavior of $B_4$ and have provided the universal values of the Binder cumulant, exactly at the critical temperature at the limit of very large volumes. These new FRG results are in a very good agreement with spin-model lattice simulations \cite{Cucchieri:2002hu, Kanaya:1994qe}.

Furthermore, we have investigated the influence of a finite symmetry breaking and finite volumes on the behavior of $B_4(T_C)$. We have found that in our calculations the finite-mass corrections are small in comparison to the corrections caused by the finite volumes. We have also shown that non-universal finite-size corrections can be described very well by taking only the leading order corrections, i.e., by contributions associated with the first irrelevant operator in the RG-flow into account. We have seen that for both models such corrections are smaller than $8\%$ for volumes with $L=10 \text{ fm}$, which are typical for lattice QCD simulations. This observation can be used in the analysis of $N_f=2$ or $N_f=(2+1)$ lattice QCD results: In Fig.~\ref{fig:BinderCumulantDifferentVolumesHVariationAtTCO2ANDO4RescaledAllData} we have illustrated that even in the presence of finite-size corrections arising in our calculations for $L=10 \text{ fm}$, O(2)- and O(4)-models can still be distinguished in an unambiguous manner. Therefore, if we assume lattice QCD to fall into either the O(2) or the O(4) universality class, then the intervals into which the values for $B_4(T_C)$ measured in lattice QCD simulations are expected to fall are clearly different for O(2) and O(4) universality classes. So, even for simulations with $L \sim 10 \text{ fm}$ we can use the universal value of the Binder cumulant exactly at $T=T_C$ to determine the universality class of the chiral transition in $N_f=2$ or $N_f=(2+1)$ lattice QCD. However, in order to apply this method, lattice results should be in the finite-size scaling region. Also the non-universal finite-size corrections in lattice QCD can be potentially larger than in our approach.

In conclusion, we have performed a new finite-size scaling analysis of the critical behavior in O(2)- and O(4)-models. Thereby we have answered the open question whether the finite-size scaling regions for O(2)- and O(4)-models differ or not. The difference we have observed is too small to be used in the scaling analysis of lattice QCD. In our investigation of the Binder cumulant in the context of the FRG, we have found that $B_4(T=T_C)$ seems to be an appropriate tool to determine the nature of the chiral transition in $N_f=2$ or $N_f=(2+1)$ lattice QCD simulations. We hope that these results will contribute fruitfully to the scaling analysis of lattice QCD.

\appendix

\section{Binder cumulants from the effective action}\label{BinderCumulant}

According to Eq.~\eqref{eq:BinderDefinition}, the $4^{\text{th}}$-order Binder Cumulant for the O(2)-model is given by
\begin{align}
B_4=\frac{\langle \sigma^4 \rangle + \langle \pi^4 \rangle + 2\langle \sigma^2 \pi^2 \rangle}{\langle \sigma^2 \rangle^2 + \langle \pi^2 \rangle^2 +2\langle \sigma^2 \rangle \langle \pi^2 \rangle}\text{ .}
\end{align}
Correlations appearing in this expression can be calculated as follows: For a model with no spatial dependence an $n$-point correlation function is defined as
\begin{equation}
\langle \rho^n \rangle = \frac{1}{V^n} \frac{1}{Z} \frac{\partial^n}{\partial H_{\rho}^n} Z \text{ ,}
\end{equation}
where $Z$ is the generating functional and $\rho= \{ \sigma, \pi \}$ is a generalized field. $\langle \rho^n \rangle$ contains connected and disconnected parts. The disconnected part is given by a sum of products of $m$-point correlation functions with $m < n$. Thereby, all possible combinations with $\sum \limits_{i} m_i=n$ appear and are multiplied with appropriate combinatorial factors. The connected part of a $n$-point correlation function is given by a $n^{\text{th}}$-derivative of the generating functional for connected diagrams $W=\log Z$:
\begin{equation}
\langle \rho^n \rangle_{\text{conn.}}= \frac{1}{V^n}\frac{\partial^n W}{\partial H_{\rho}^n}\text{ .}
\end{equation}
The first derivative of $W$ is given by the expectation value of $\rho$. In following we call this quantity the classical field.
\begin{equation}
	\rho_{ cl}= \langle \rho \rangle =M_{\rho}= \frac{1}{V} \frac{\partial W}{\partial H_{\rho}}\text{ .}
	\label{eq:ExpectationValueDefinition}
\end{equation}
The second derivative of $W$ is coupled to susceptibility
\begin{equation}
\frac{\partial^2 W}{\partial H_{\rho}^2}=\frac{\chi_{\rho}}{V}=\frac{1}{V m_{\rho}^2} \text{ .}
\end{equation}

This quantity is also the dressed propagator and is connected to the inverse second derivative of the effective action $\Gamma$ with respect to the classical field. In our calculations  $\Gamma=VU$ and this statement takes the form
\begin{equation}
\frac{\partial^2 W}{\partial H_{\rho_1} \partial H_{\rho_2}} =V^2 D_{\rho_1 \rho_2} = V \Big(\frac{\partial^2 U}{\partial \rho_{1,\text{cl}} \partial \rho_{2,\text{cl}}}\Big)^{-1}\text{ ,}
\label{eq:CentralEquation}
\end{equation}
where $D_{\rho_1 \rho_2}$ is a $2 \times 2$ matrix.

All higher derivatives of $W$ can be calculated using iterative application of the operator
\begin{align}
\begin{split}
\frac{\partial}{\partial H_{\rho_3}}&=\sum_{\rho_4} \frac{\partial \rho_{4,\text{cl}}}{\partial H_{\rho_3}} \frac{\partial}{\partial \rho_{4,\text{cl}}} \\ &= \sum_{\rho_4} \Big(\frac{\partial^2 U}{\partial \rho_{3,\text{cl}} \partial \rho_{4,\text{cl}}}\Big)^{-1} \frac{\partial}{\partial \rho_{4,\text{cl}}}\text{ ,}
\end{split}
\end{align}
on the Eq.~\eqref{eq:CentralEquation}.

We represent our results for the O(2)-model in terms of Feynman diagrams defined as follows: A general dressed $n$-point vertex in our theory is given by
\begin{align}
-V\frac{\partial^n U}{\partial \phi_{1, \text{cl}} \ldots \partial \phi_{n, \text{cl}}}=-\frac{\partial^n \Gamma}{\partial \phi_{1, \text{cl}} \ldots \partial \phi_{n, \text{cl}}} \text{ ,}
\end{align}
and is denoted by empty circle. A general dressed static propagator is
\begin{align}
D_{\rho_1 \rho_2}=\frac{1}{V m_{\rho_1 \rho_2}^2}\text{ .}
\end{align}
We denote it by line with a full circle. For $\sigma$ we use continuous and for $\pi$ dashed line. The non-vanishing expectation value of $\sigma$ is represented by
\begin{align}
\langle \sigma \rangle = M =
				\begin{aligned}
					\vspace{2cm}
					\includegraphics[width=0.04\textwidth]{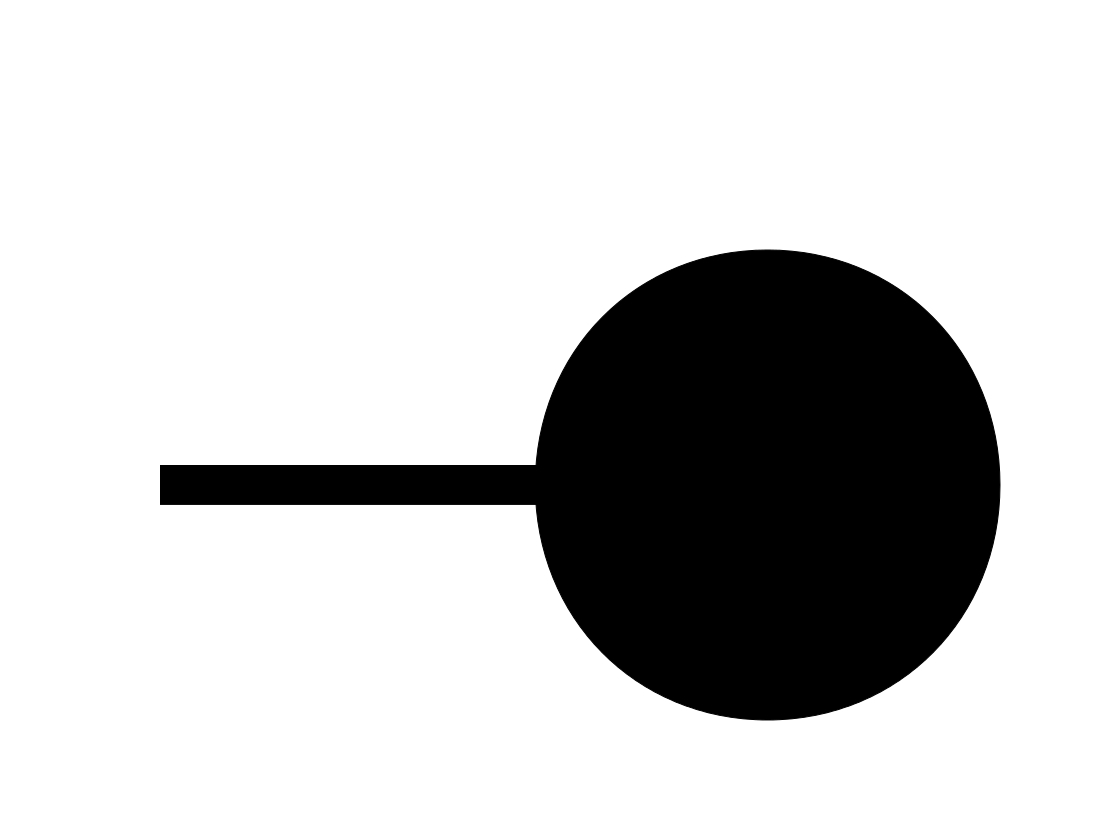}
				\end{aligned} \vspace{-1cm} \text{ .}
\end{align}
The correlation functions calculated using FRG in LPA, which are relevant for the $B_4$ of the O(2)-symmetric model, are given by
\begin{align}
\begin{split}
\langle \sigma^2 \rangle &= 
				\begin{aligned}
					\vspace{2cm}
					\includegraphics[width=0.08\textwidth]{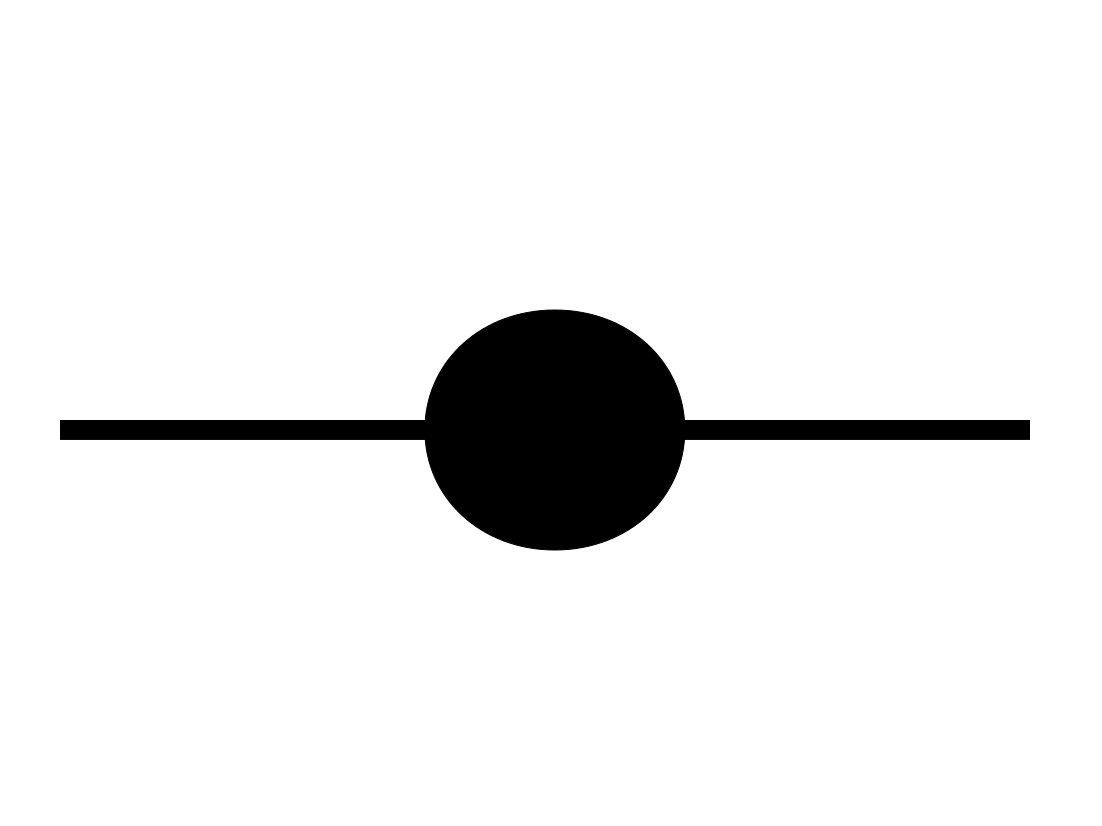}
				\end{aligned} \vspace{-1cm} +
				\begin{aligned}
					\vspace{2cm}
					\includegraphics[width=0.08\textwidth]{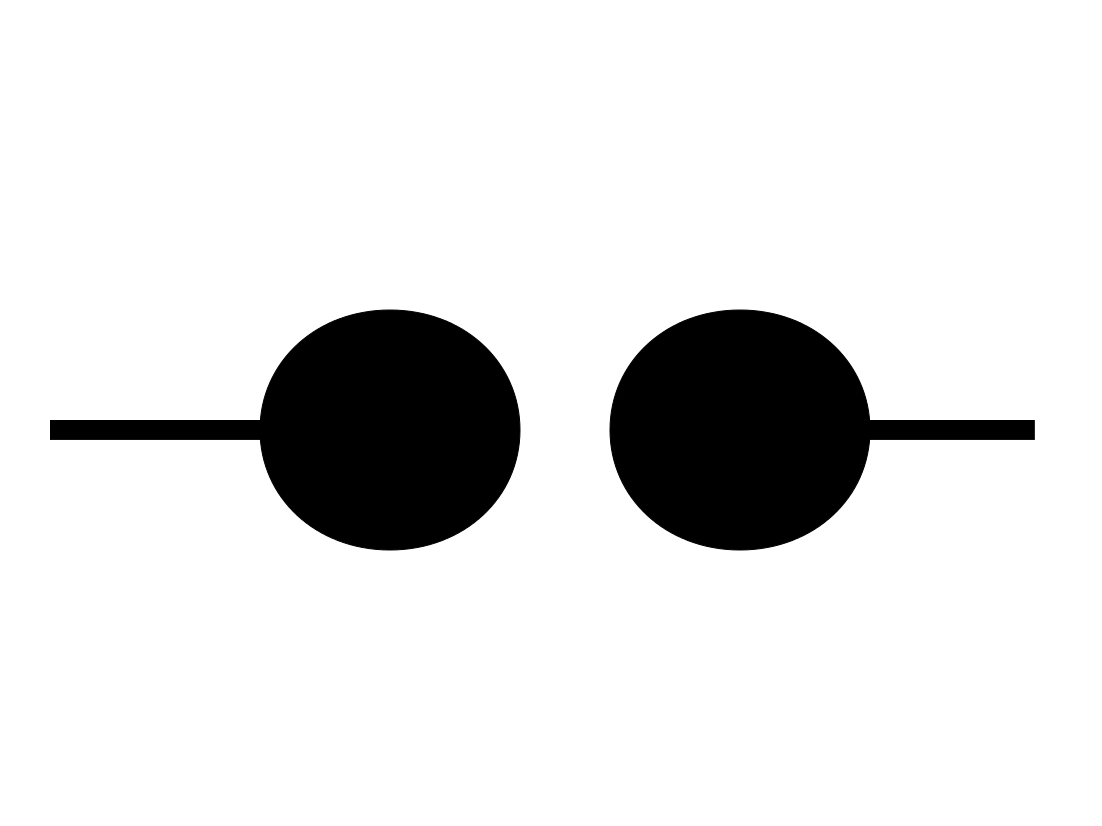}
				\end{aligned} \vspace{-1cm}\text{ ,}
\end{split}
\end{align}
\begin{align}
\begin{split}
\langle \sigma^4 \rangle &=\begin{aligned}
					\vspace{2cm}
					\includegraphics[width=0.08\textwidth]{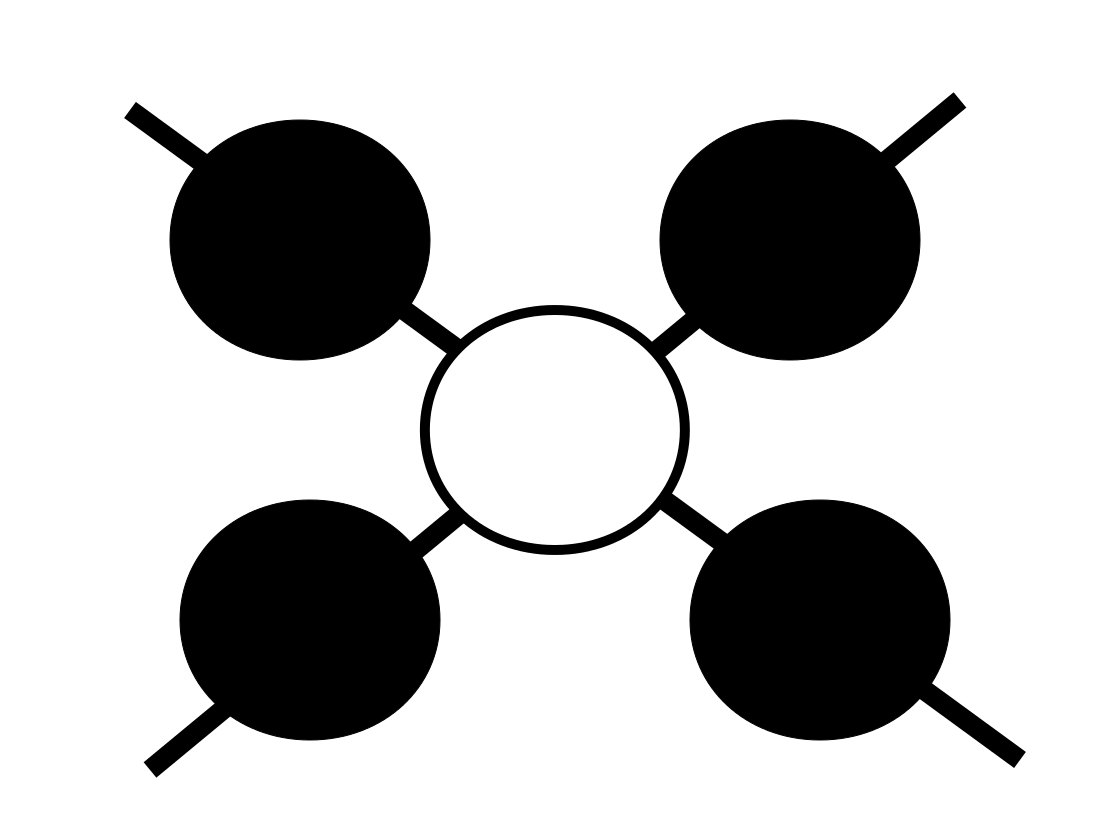}
				\end{aligned} \vspace{-1cm} + 3
				\begin{aligned}
					\vspace{2cm}
					\includegraphics[width=0.08\textwidth]{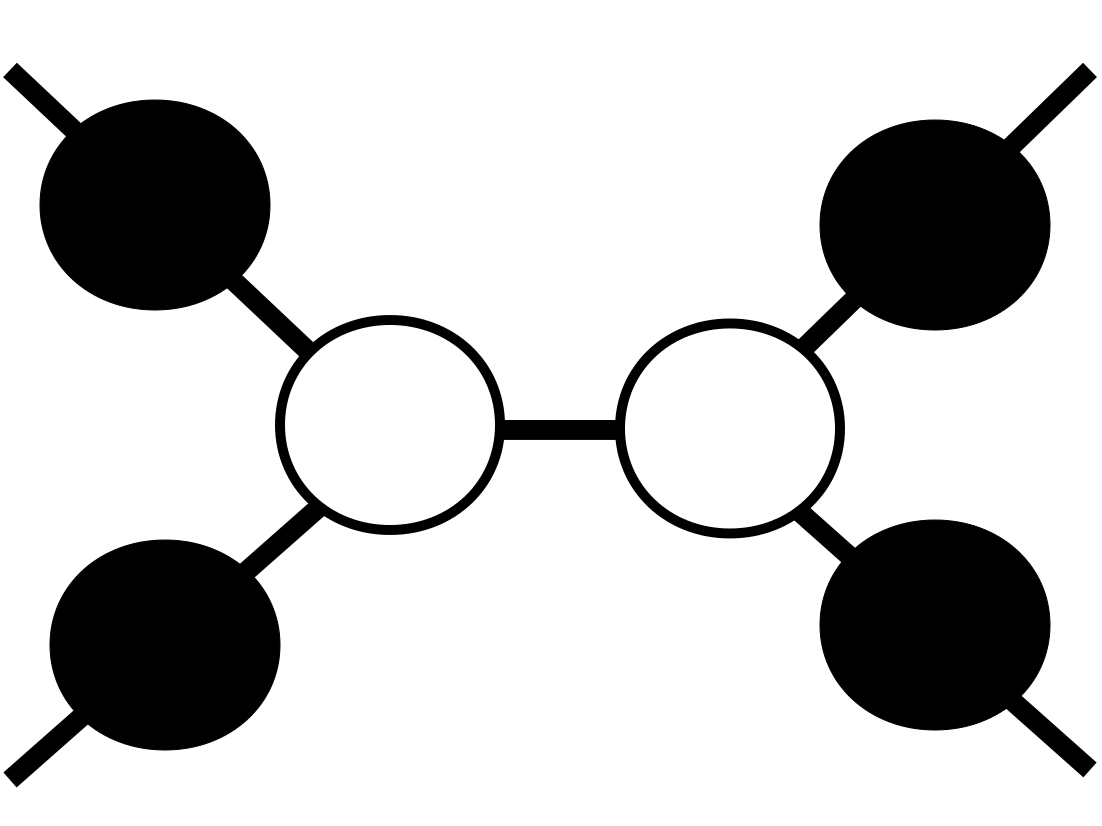}
				\end{aligned} \vspace{-1cm} +3 
				\begin{aligned}
					\vspace{2cm}
					\includegraphics[width=0.08\textwidth]{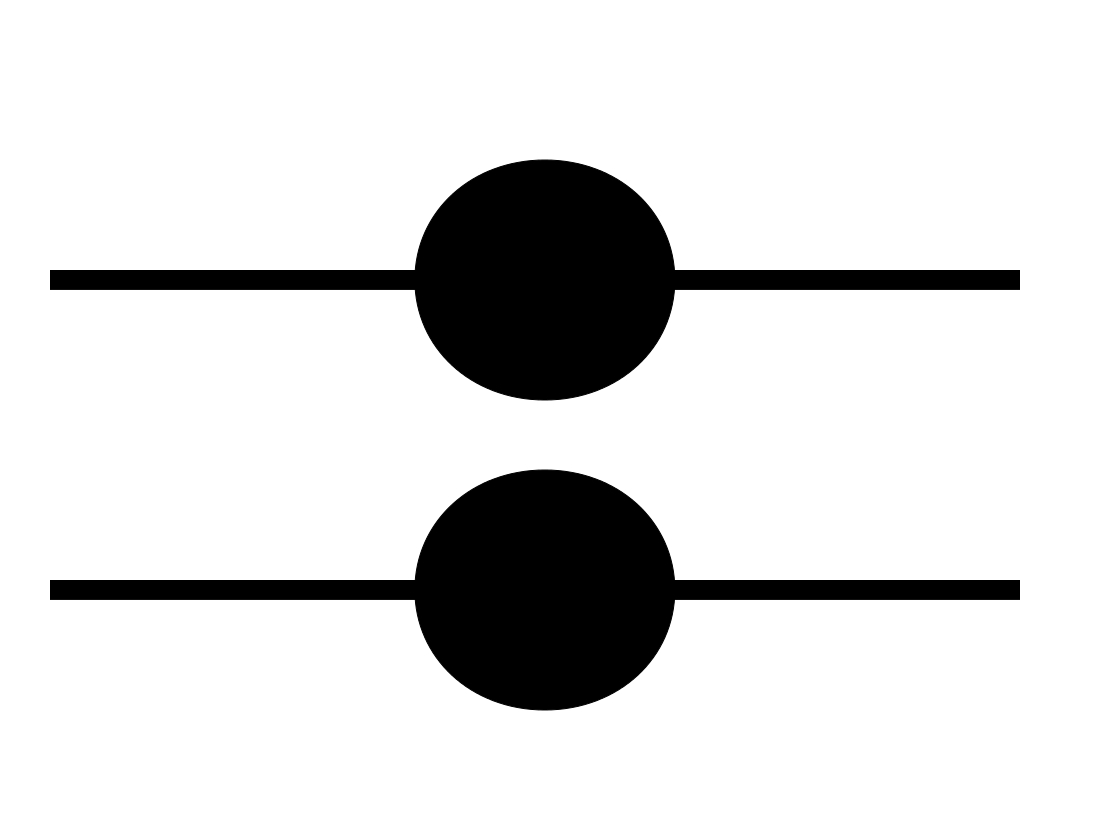}
				\end{aligned} \vspace{-1cm}\\ & \enspace \enspace + 4
				\begin{aligned}
					\vspace{2cm}
					\includegraphics[width=0.08\textwidth]{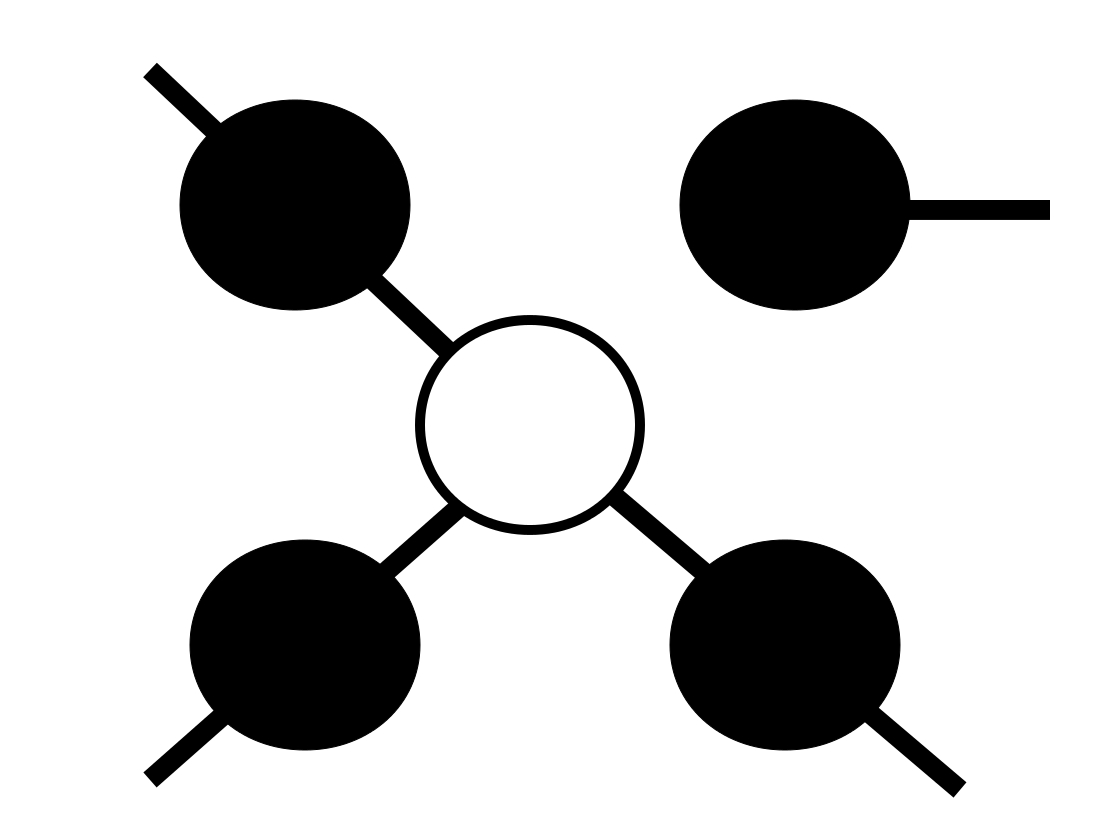}
				\end{aligned} \vspace{-1cm} + 6
				\begin{aligned}
					\vspace{2cm}
					\includegraphics[width=0.08\textwidth]{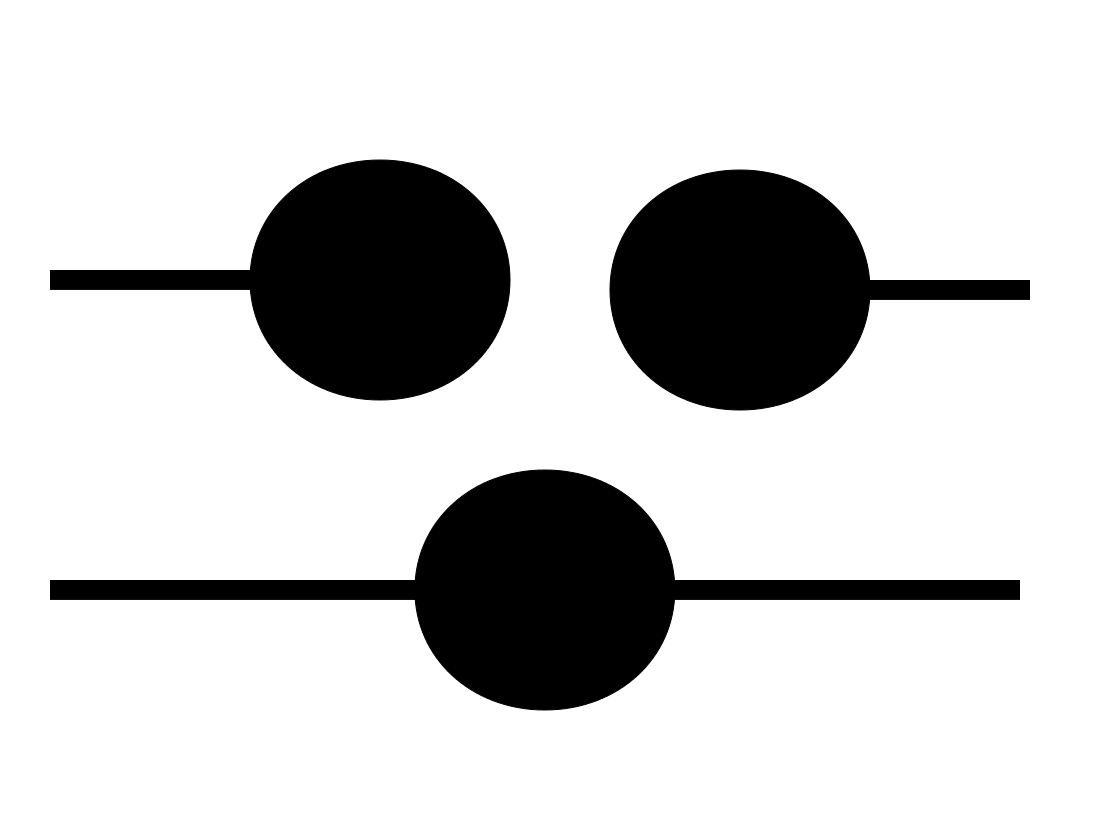}
				\end{aligned} \vspace{-1cm} + 
				\begin{aligned}
					\vspace{2cm}
					\includegraphics[width=0.08\textwidth]{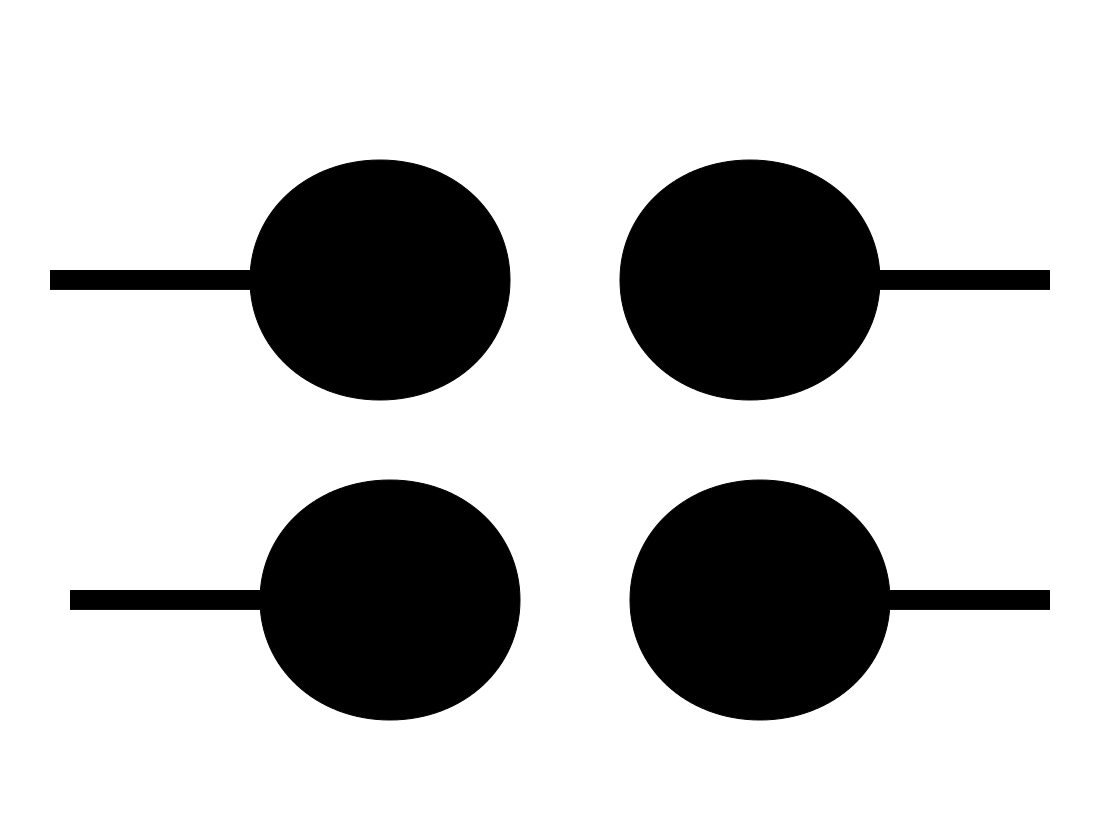}
				\end{aligned} \vspace{-1cm}\text{ ,}
\end{split}
\end{align}
\begin{align}
\begin{split}
\langle \pi^2 \rangle &=
				\begin{aligned}
					\vspace{2cm}
					\includegraphics[width=0.08\textwidth]{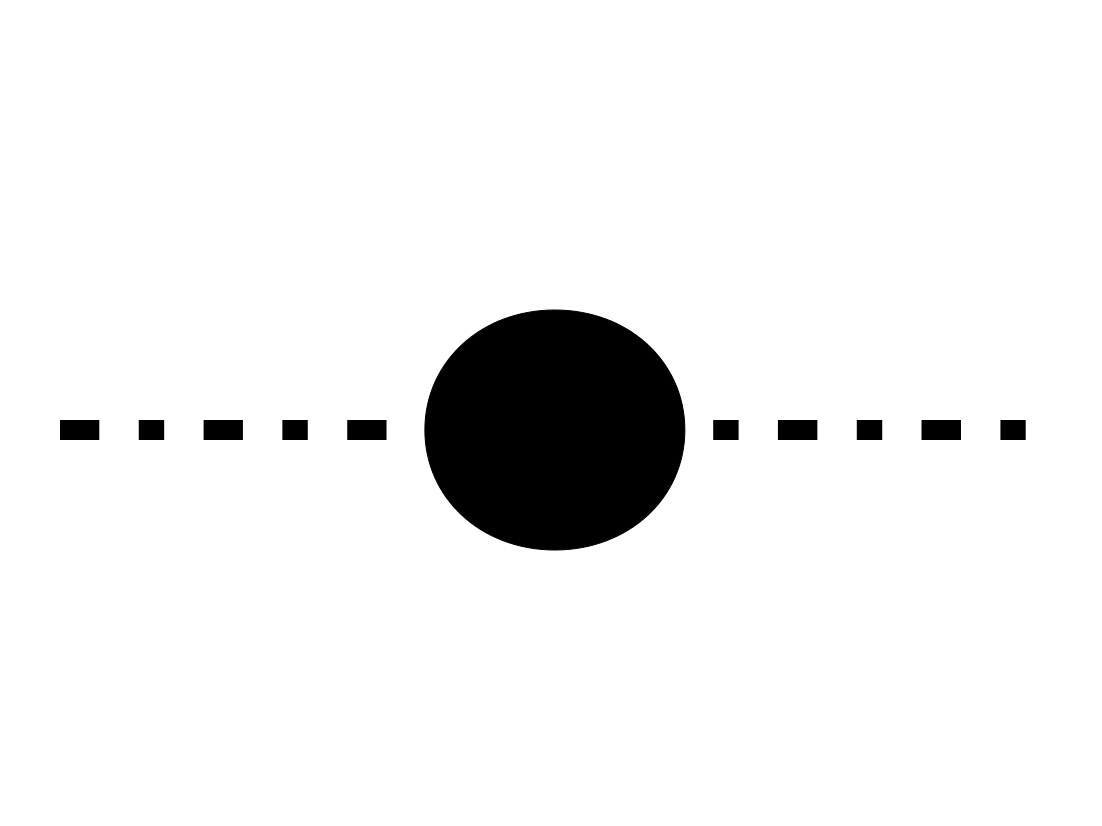}
				\end{aligned} \vspace{-1cm}\text{ ,}
\end{split}
\end{align}
\begin{align}
\begin{split}
\langle \pi^4 \rangle &=
				\begin{aligned}
					\vspace{2cm}
					\includegraphics[width=0.08\textwidth]{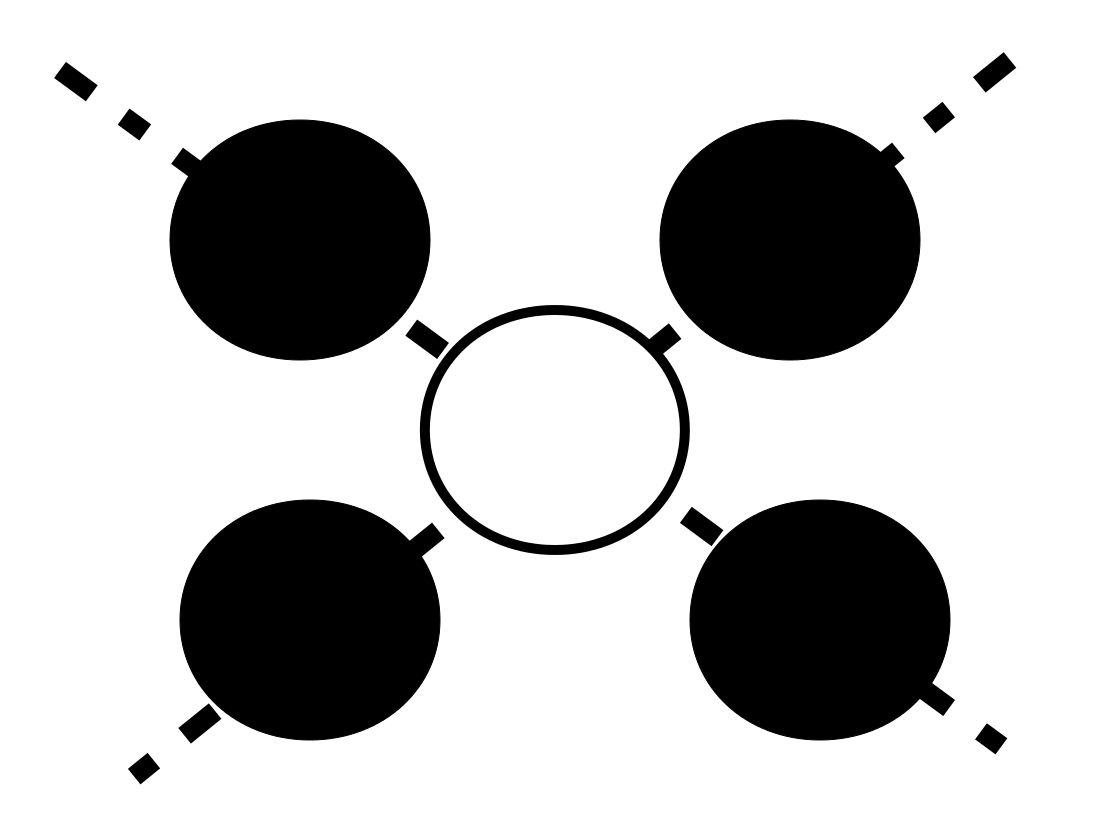}
				\end{aligned} \vspace{-1cm} + 3
				\begin{aligned}
					\vspace{2cm}
					\includegraphics[width=0.08\textwidth]{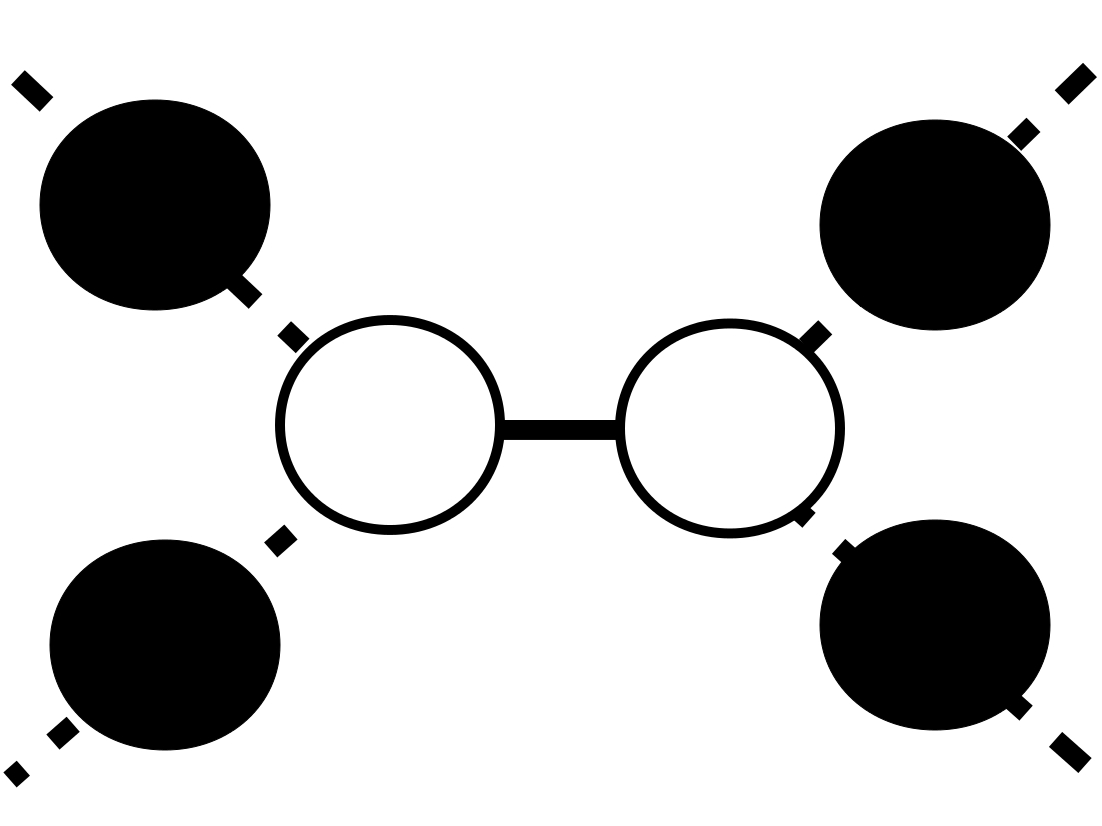}
				\end{aligned} \vspace{-1cm} + 3
				\begin{aligned}
					\vspace{2cm}
					\includegraphics[width=0.08\textwidth]{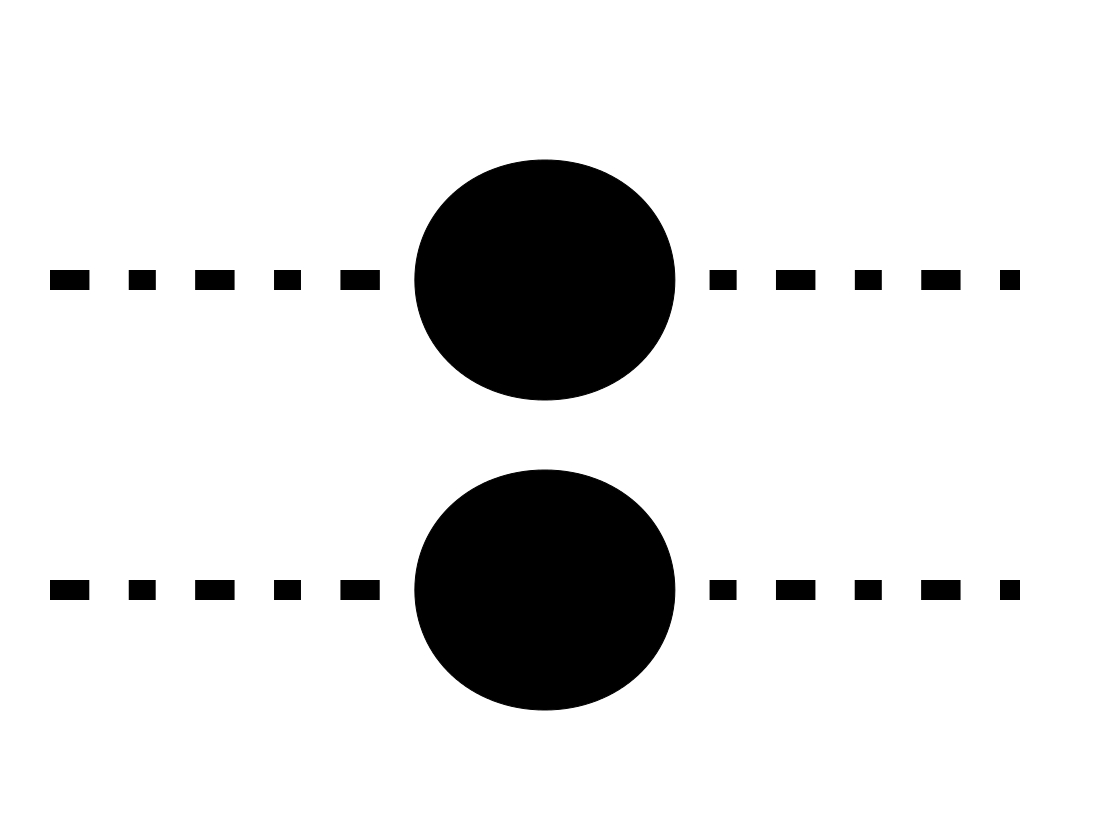}
				\end{aligned} \vspace{-1cm}\text{ ,}
\end{split}
\end{align}
\begin{align}
\begin{split}
\langle \pi^2 \sigma^2 \rangle &=\begin{aligned}
					\vspace{2cm}
					\includegraphics[width=0.08\textwidth]{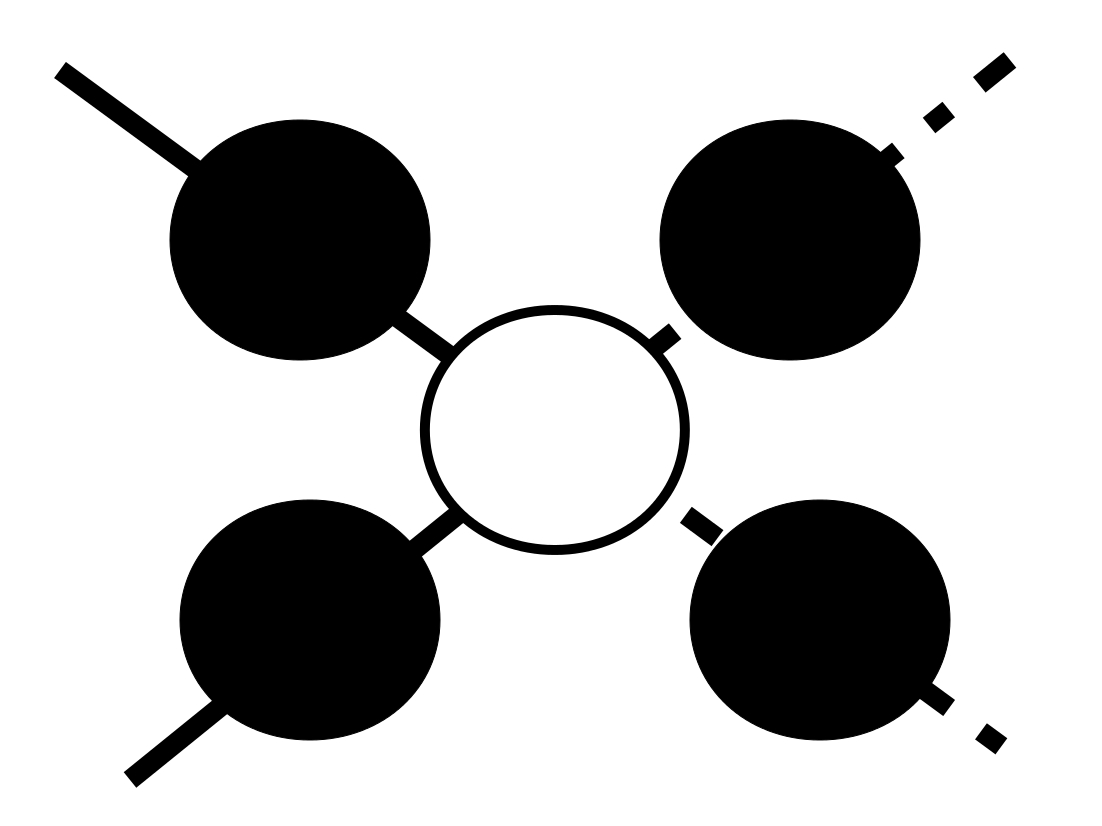}
				\end{aligned} \vspace{-1cm} + 
				\begin{aligned}
					\vspace{2cm}
					\includegraphics[width=0.08\textwidth]{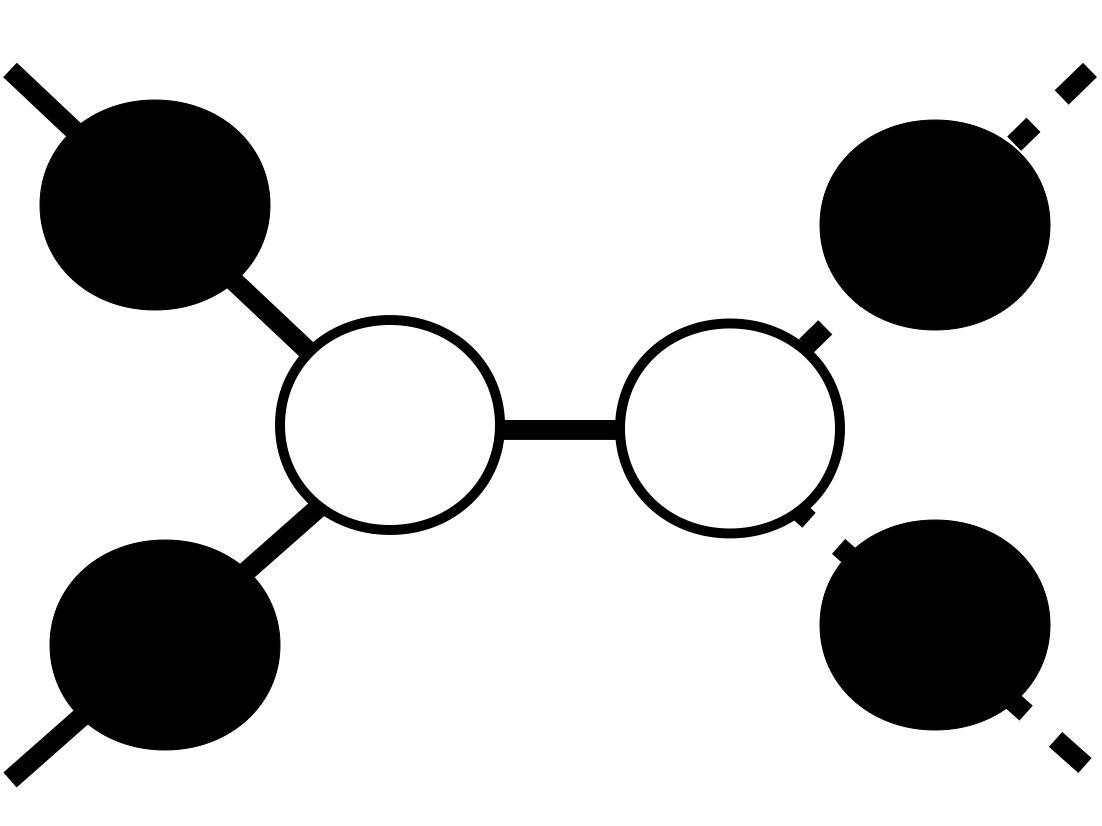}
				\end{aligned} \vspace{-1cm} +2
				\begin{aligned}
					\vspace{2cm}
					\includegraphics[width=0.08\textwidth]{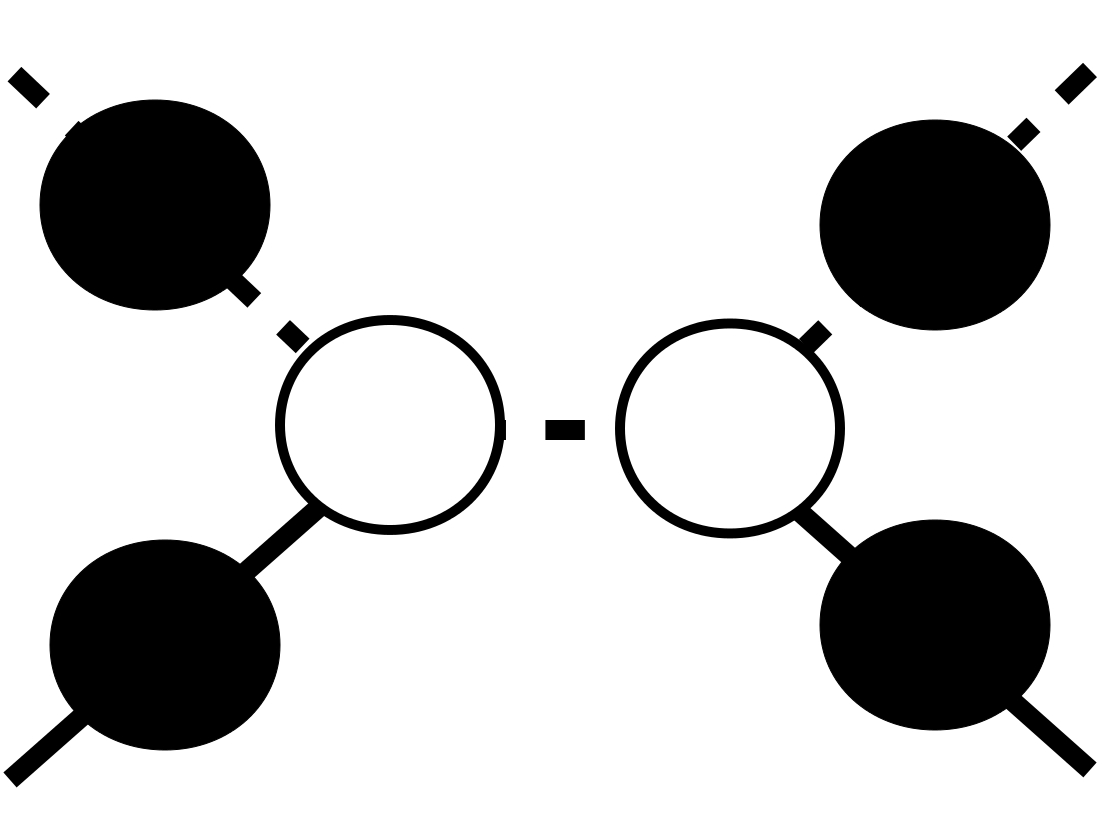}
				\end{aligned} \vspace{-1cm}\\ & \enspace \enspace + 2
				\begin{aligned}
					\vspace{2cm}
					\includegraphics[width=0.08\textwidth]{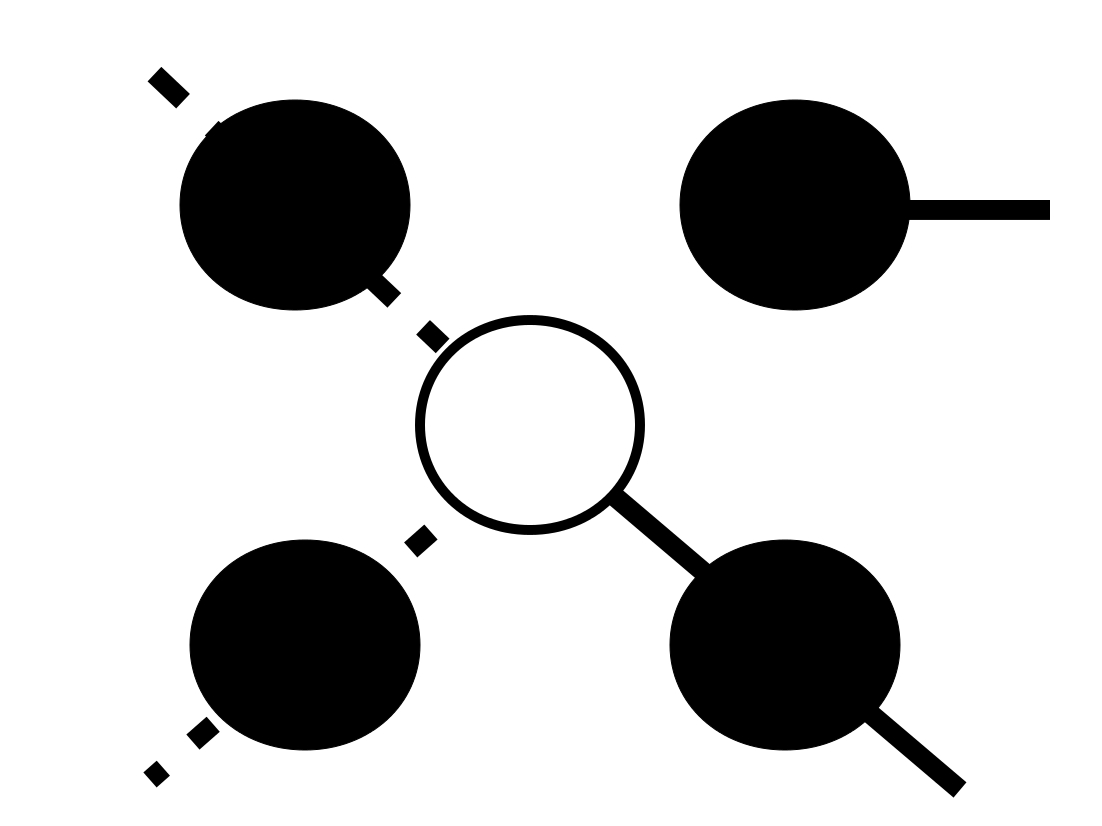}
				\end{aligned} \vspace{-1cm} + 
				\begin{aligned}
					\vspace{2cm}
					\includegraphics[width=0.08\textwidth]{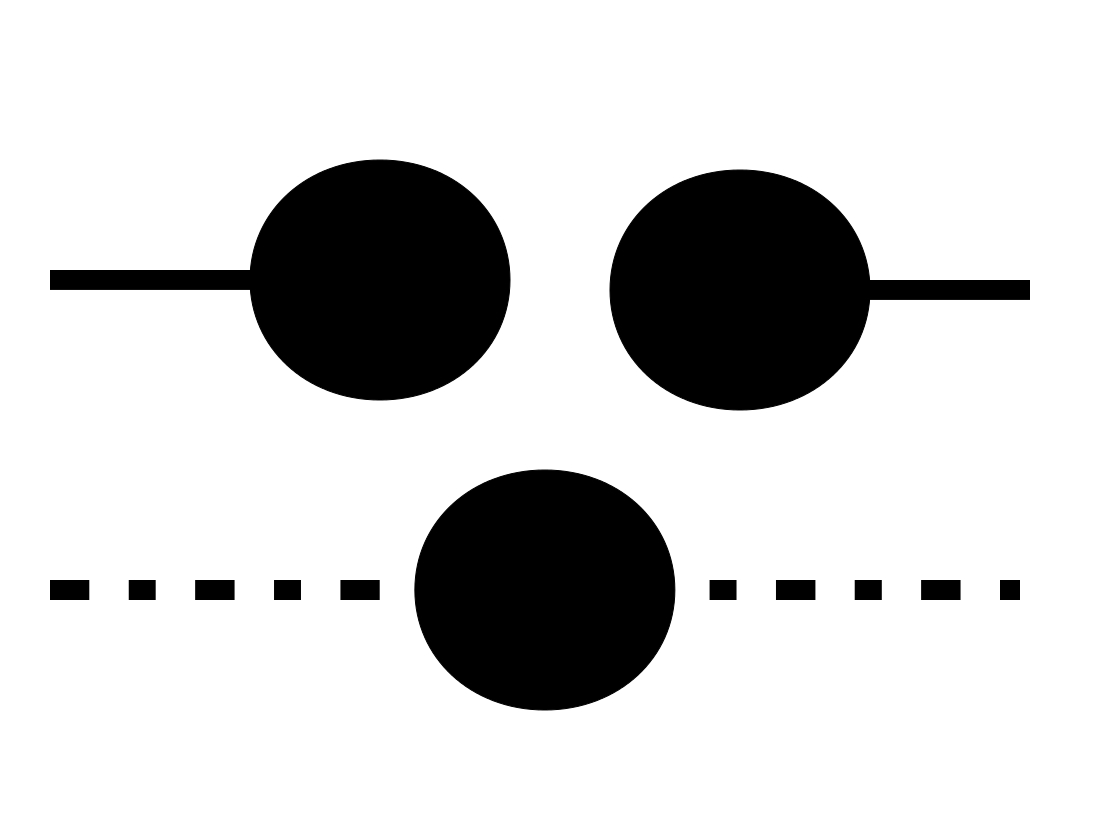}
				\end{aligned} \vspace{-1cm} + 
				\begin{aligned}
					\vspace{2cm}
					\includegraphics[width=0.08\textwidth]{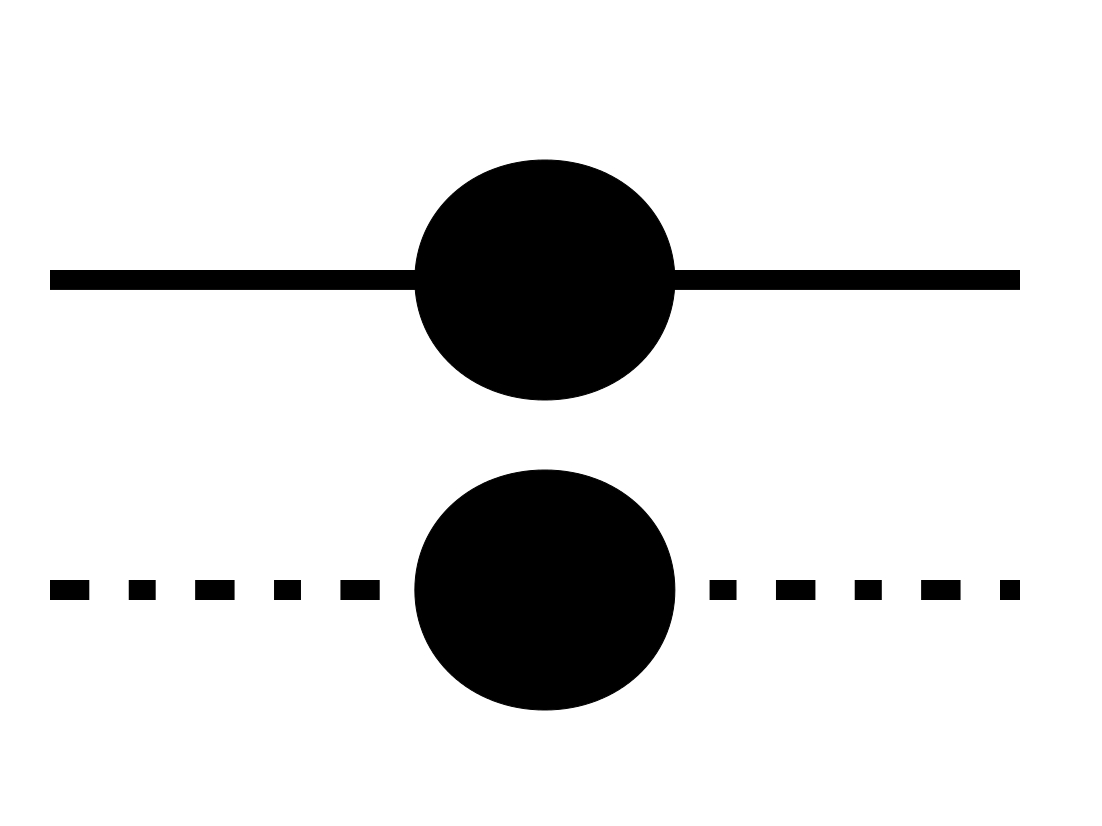}
				\end{aligned} \vspace{-1cm}\text{ .}
\end{split}
\end{align}

From these expressions we see that in the thermodynamic limit and in the phase with broken symmetry the Binder cumulant approaches 1 in the leading order. In the limit of restored symmetry, the order parameter $M$ vanishes. Therefore, we have to look at next-to-leading order terms in $1/V$. We also have to keep in mind that for $T \gg T_C$ masses of $\pi$ and $\sigma$ become equal. So, we obtain $B_4=2$:

In the case of O(4)-model calculations are very similar but now we have three different pion fields. However, they are completely equivalent. Therefore, all correlation functions with no mixing of different pion fields will provide exactly the same results and can be calculated using the same expressions as for O(2)-model.

In correlations with mixing of pions, two different pion fields are involved. However any combination of different components will lead to one and the same result
\begin{align}
\langle \pi_i^2 \pi_j^2 \rangle = \langle \pi_1^2 \pi_2^2 \rangle \text{ ,}
\end{align}
with $i \neq j$. Using this, we get the following expression for the Binder cumulant of the O(4)-model:
\begin{align}
B_4=\frac{\langle \sigma^4 \rangle + 3\langle \pi^4 \rangle + 6\langle \pi^2 \sigma^2 \rangle + 6\langle \pi_1^2 \pi_2^2 \rangle}{\langle \sigma^2 \rangle^2 + 9 \langle \pi^2 \rangle^2 + 6 \langle \sigma^2 \rangle \langle \pi^2 \rangle}\text{ .}
\end{align}
The only correlation in the expression above, which we have to evaluate in addition, is $\langle \pi_1^2 \pi_2^2 \rangle$:
\begin{align}
\begin{split}
\langle \pi_{1}^2 \pi_{2}^2 \rangle &=\begin{aligned}
					\vspace{2cm}
					\includegraphics[width=0.1\textwidth]{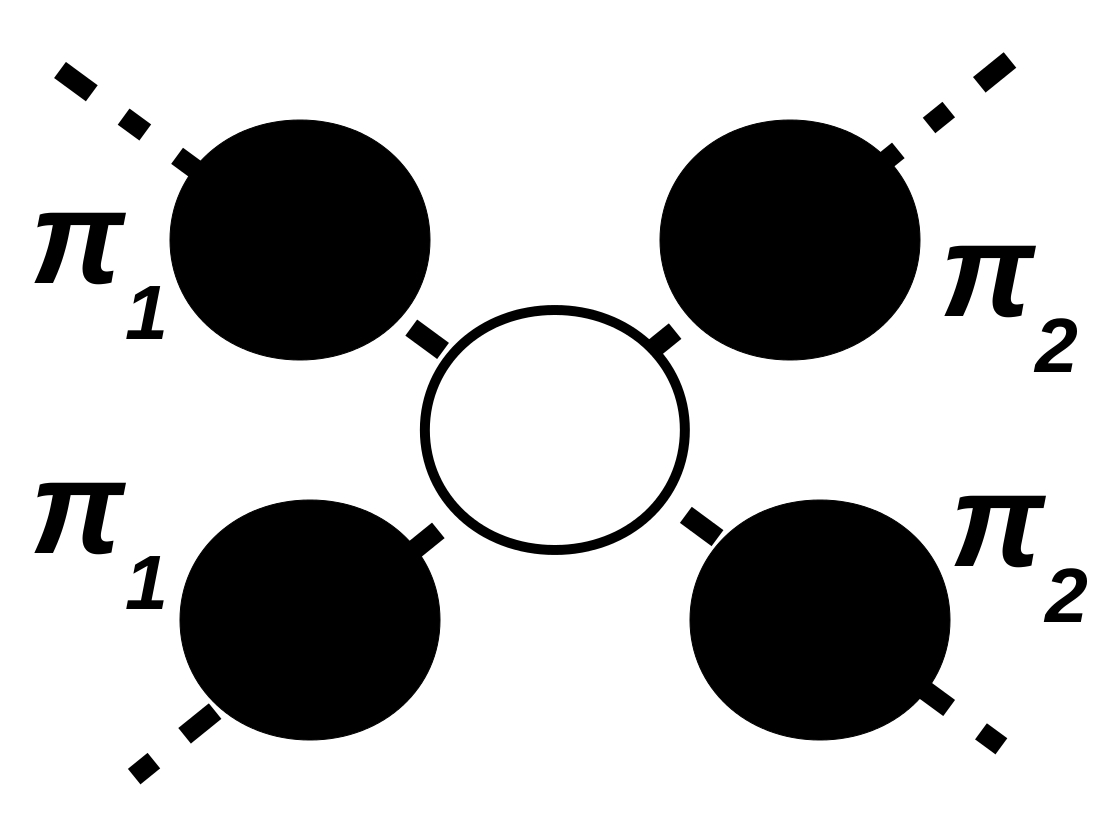}
				\end{aligned} \vspace{-1cm} + 
				\begin{aligned}
					\vspace{2cm}
					\includegraphics[width=0.1\textwidth]{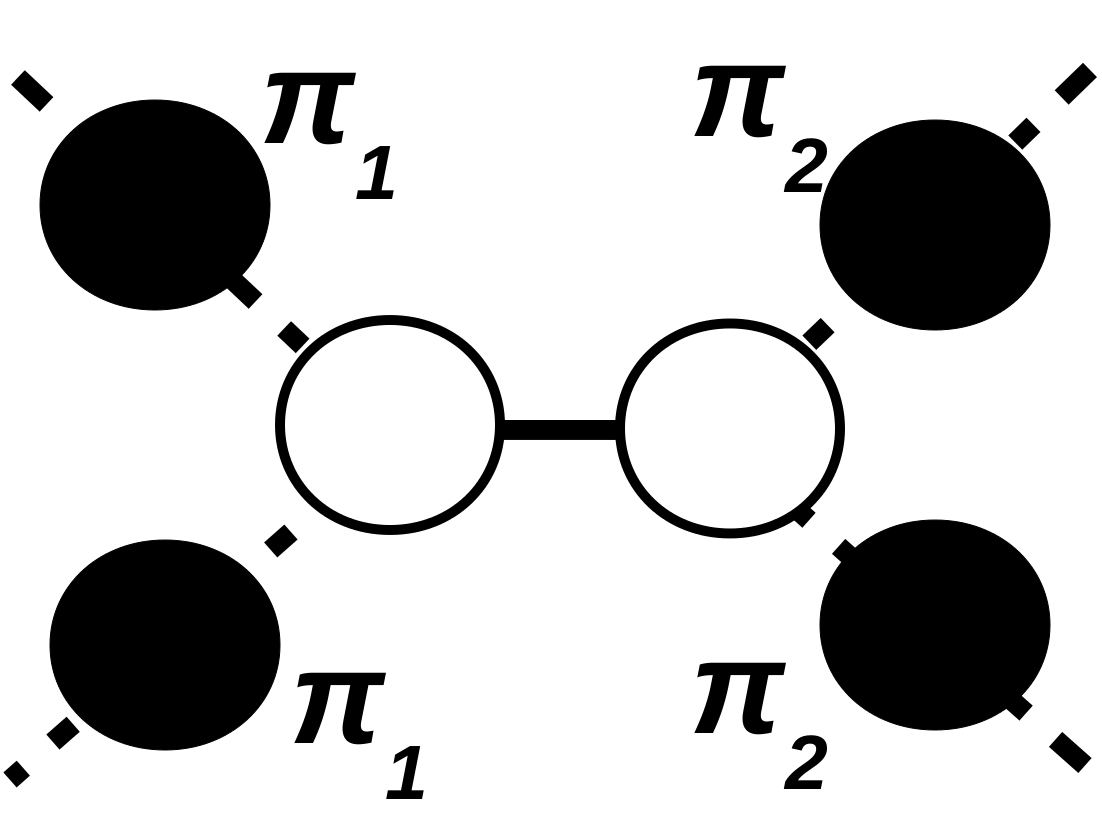}
				\end{aligned} \vspace{-1cm}+ 
				\begin{aligned}
					\vspace{2cm}
					\includegraphics[width=0.1\textwidth]{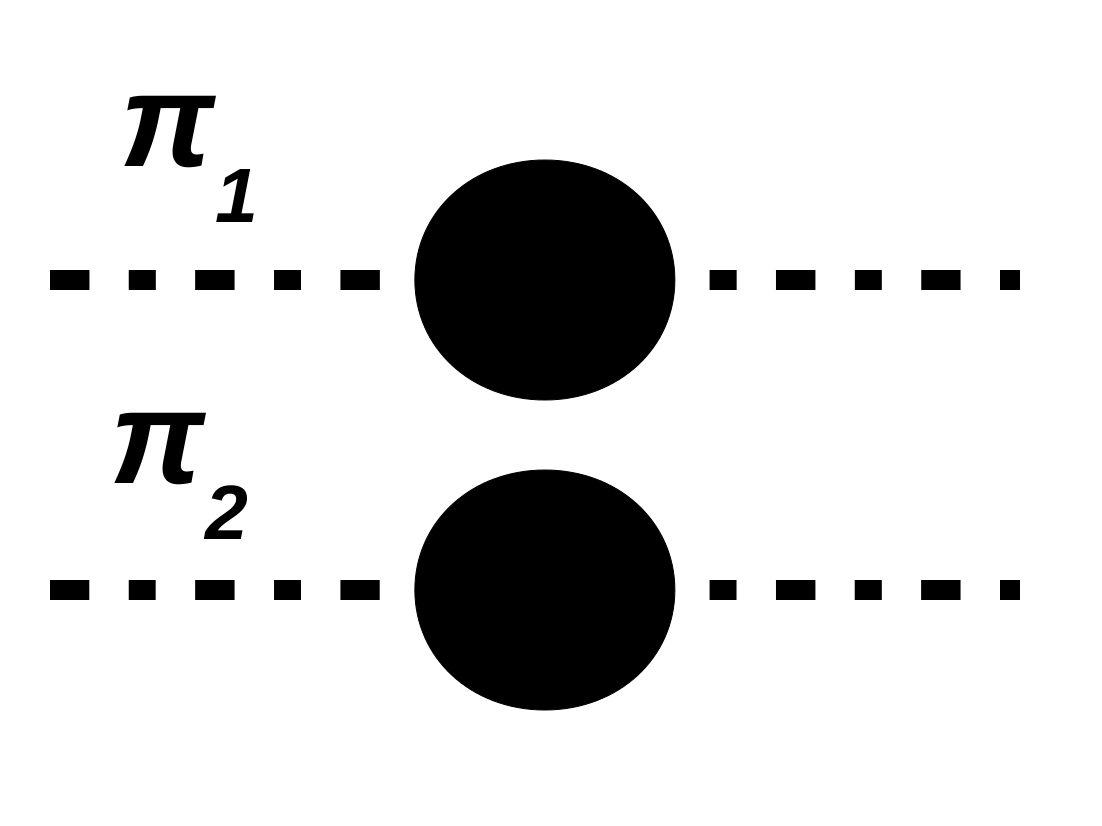}
				\end{aligned} \vspace{-1cm}\text{ .}
\end{split}
\end{align}

Using this result we have found that limiting behavior of $B_4$ for the O(4)-model in the thermodynamic limit is given  by 1 for the low-temperature and by 3/2 for the high-temperature phase.

\end{document}